\documentclass[prx,aps,amsmath,amssymb,nofootinbib,epsfig,showpacs,twocolumn,eqsecnum,superscriptaddress]{revtex4-1}

\usepackage{amsmath,cleveref,amssymb,amsfonts,graphicx,multirow,color,bm,textcomp,mathtools}
\usepackage{paralist}
\usepackage{srcltx}
\usepackage[all,cmtip]{xy}
\usepackage{mathrsfs}
\usepackage{srcltx,soul,subfigure}
\usepackage{threeparttable,dsfont,bbm}

\crefname{equation}{Eq.}{Eqs.}

\newcommand{\bra}[1]{\langle #1 |}
\newcommand{\ket}[1]{| #1 \rangle}
\newcommand{\braket}[2]{\left \langle #1 | #2 \right\rangle}

\newcommand{\bee}{\begin{eqnarray}}
\newcommand{\ee}{\end{eqnarray}}
\newcommand{\bma}{\begin{pmatrix}}
\newcommand{\ema}{\end{pmatrix}}
\newcommand{\balig}{\begin{align}}
\newcommand{\ealig}{\end{align}}

\newcommand{\ba}{\begin{align}}
\newcommand{\ea}{\end{align}}

\newcommand{\ignore}[1]{}

\newcommand{\six}{\sigma_x}
\newcommand{\siy}{\sigma_y}
\newcommand{\siz}{\sigma_z}

\newcommand{\bI}{\mathbbm{1}}

\def\sgn{\mathop{\textrm{sgn}}}

\newcommand{\bk}{{\bf k}}
\newcommand{\bs}{{\bf s}}
\newcommand{\bR}{{\bf R}}

\usepackage{dcolumn}
\newcolumntype{C}[1]{>{\centering\let\newline\\\arraybackslash\hspace{0pt}}m{#1}}

\begin{document}

%\title{Topological line nodes in Ca$_3$P$_2$ and other semi-metals}
%\title{Topological semi-metals with line nodes and nearly flat surface states} 
\title{Topological semi-metals with line nodes and drumhead surface states}

\author{Y.-H. Chan}
%\email{yanghao@umich.edu}
\affiliation{Institute of Atomic and Molecular Sciences, Academia Sinica, Taipei 10617, Taiwan}

\author{Ching-Kai Chiu}
%\email{chiu7@phas.ubc.ca}
\affiliation{Department of Physics and Astronomy, University of British Columbia, Vancouver, BC, Canada V6T 1Z1} 
\affiliation{Quantum Matter Institute, University of British Columbia, Vancouver BC, Canada V6T 1Z4}
\affiliation{Condensed Matter Theory Center, Department of Physics, University of Maryland, College Park, MD 20742, USA}

\author{M. Y. Chou}
\affiliation{Institute of Atomic and Molecular Sciences, Academia Sinica, Taipei 10617, Taiwan}
\affiliation{School of Physics, Georgia Institute of Technology, Atlanta, GA 30332, USA}

\author{Andreas P. Schnyder}
\email{a.schnyder@fkf.mpg.de}
\affiliation{Max-Planck-Institut f\"ur Festk\"orperforschung, Heisenbergstrasse 1, D-70569 Stuttgart, Germany}

\begin{abstract}
In an ordinary three-dimensional metal the Fermi surface forms a two-dimensional closed
sheet separating the filled from the empty states. Topological semimetals, on the other hand,
can exhibit protected one-dimensional Fermi lines or zero-dimensional Fermi points,
which arise due to an intricate interplay between symmetry and topology of the electronic wavefunctions.
Here, we study how reflection symmetry, time-reversal symmetry, SU(2) spin-rotation symmetry,
and  inversion symmetry lead
to the topological protection of line nodes in three-dimensional semi-metals. We obtain
the %$\mathbbm{Z}$- and $\mathbbm{Z}_2$-type
crystalline invariants that guarantee the stability
of the line nodes in the bulk and show that a quantized Berry phase leads to the appearance of protected surfaces states
with a nearly flat dispersion.
By deriving a relation between the crystalline invariants and the Berry phase, we
establish a direct connection between the stability of the line nodes and the topological surface states.
As a representative example of a topological semimetal with line nodes, we consider
Ca$_3$P$_2$ and discuss the topological properties of its Fermi line 
in terms of a low-energy effective theory and a  tight-binding model, derived from ab initio DFT calculations. 
Due to the bulk-boundary correspondence, Ca$_3$P$_2$ displays
nearly dispersionless surface states, which take the shape of a drumhead.
 These surface states could potentially give rise to novel topological response
phenomena and provide an avenue for exotic correlation physics at the surface.
%Comment on experimental implications.
% highly unusual ring of Dirac nodes at the Fermi level. 
\end{abstract}

\date{\rm\today}

\maketitle

%%%%%%%%%%%%%
%%%%%%%%%%%%%%%%%%

\section{Introduction}

The study of band structure topology of insulating and semi-metallic materials has become an increasingly important topic in modern condensed matter physics~\cite{hasan:rmp,qi:rmp,chiu_review15,Schnyder2008,Ryu2010ten}.
The discovery of spin-orbit induced topological insulators	
has revealed that a non-trivial 
momentum-space topology of the electronic bands
can give rise to new  states of matter with exotic surface
states~\cite{Hsieh:2008fk,hsiehNature2009,Xia:2009uq,andoJPSJreview13,HasanMoore2011,reviewBookHasan} and highly unusual magneto-transport properties~\cite{Qi2008sf,QiHughesZhang2007,garateFranzPRB10}.
Recently, due to the experimental detection of arc surface states in Weyl semi-metals~\cite{XuHasanScience15},
considerable attention 
has focused on the investigation 
of topological semi-metals~\cite{matsuuraNJP13,ZhaoWangPRL13,ChiuSchnyder14,FangFu_PRB15,BurkovBalentsPRB11,ZhaoWangPRB14,Ashvin_Weyl_review,krempaBallentsAnnuRev2014,Hosur_Weyl_develop,
Xu_Weyl_2015_first,Lu_photonic_Weyl_2015,Weyl_discovery_TaAs,gaoZhangArxiv15,kim_kane_rappe_PRL_15,wengKawazonePRB15,chen_fan_zhang_arXiv_15}.
While in ordinary three-dimensional metals
filled and empty states are separated by
two-dimensional Fermi sheets, 
topological semi-metals can exhibit
zero-dimensional Ferm points 
or one-dimensional Fermi lines.

Classic examples of  topological semi-metals are
the Weyl and Dirac semi-metals which exhibit 
two-fold and four-fold degenerate Fermi points, respectively.
 Weyl points can occur  in 
the absence of any symmetry besides translation, whereas Dirac points are 
topologically stable only in the presence
of time-reversal symmetry together with a
crystal lattice symmetry, such as rotation or reflection.
For example in the Dirac materials 
Cd$_3$As$_2$~\cite{Yazdani_CdAs,Dirac_semimetal_Xi_Dai,neupaneDiracHasan,borisenkoPRLCd3As2,Cd3As2Chen2014,liangOngTransportCd3As2}
 and Na$_3$Bi~\cite{Liu21022014,Dai_predition_Na3Bi,xuLiuHasanArxiv13,Xu18122014,Chiu_C4_Dirac},
 the gapless property of the Dirac points is protected by a $C_4$ and $C_3$ crystal rotation symmetry, respectively. 
 Correspondingly, the stability of Weyl points is guaranteed by a Chern number, 
 while Dirac points are protected by a crystalline invariant, e.g., a mirror number~\cite{chiu_review15}.
 Due to their topological characteristics these point-node semi-metals display
 a number of exotic transport phenomena, such as negative magneto-resistance
 and chiral magnetic effect~\cite{Burkov_Weyl_electromagnetic_2012,Hosur_Weyl_develop,Lu_anomaly_Weyl_2013,Sid_anomaly_Weyl,Marcel_Weyl_response}. 
 
Probably even more interesting than semi-metals with point nodes are topological materials with line nodes,
since they support weakly dispersing
surface states that could provide an interesting platform
for exotic correlation physics~\cite{KopninVolovikPRB11,tangFu_natPhys14,graphene_edge_magnetism}. 
Moreover, these semi-metals are expected to exhibit long-range Coulomb interaction~\cite{Nodal_line_coulomb} and graphene-like Landau levels~\cite{Nodal_ring_Landau_level}. 
In nodal line semi-metals the valence and conduction bands cross along one-dimensional
lines in momentum space forming a ring-shaped Fermi line. 
From the general classification of gapless topological materials~\cite{chiu_review15} it follows, that line nodes in semi-metals are stable against gap opening only in the presence of a lattice symmetry, such as, e.g., reflection~\cite{ChiuSchnyder14,FangFu_PRB15,BurkovBalentsPRB11}. That is,
 the two bands that cross at (or near)  the Fermi level  of a nodal line semi-metal
have opposite crystal symmetry eigenvalues, which prevents hybridization. 
For example, in non-centrosymmetric PbTaSe$_2$~\cite{cava_PbTaSe2_PRB_14,bian_hasan_arXiv_15}
and TlTaSe$_2$~\cite{bian_hasan_arXiv_15b} 
the reflection about the Ta atomic planes protects the topological nodal lines.
Similarly, the band crossings  in Cu$_3$PdN~\cite{yu_Ca3PdN_arXiv}, 
 ZrSiS~\cite{schoopZrSiS_arXiv15}, and Ca$_3$P$_2$~\cite{xie_schoop_arXiv}
are protected by point group symmetries. 
Since the latter three systems are symmetric under both inversion and time reversal, their nodal rings
are four-fold degenerate, i.e., of ``Dirac type".
In contrast,  PbTaSe$_2$ and TlTaSe$_2$ lack inversion symmetry and hence exhibit
``Weyl rings", which are only two-fold degenerate. 

\begin{figure}[h!]
\begin{center}
\includegraphics[clip,width=0.95\columnwidth]{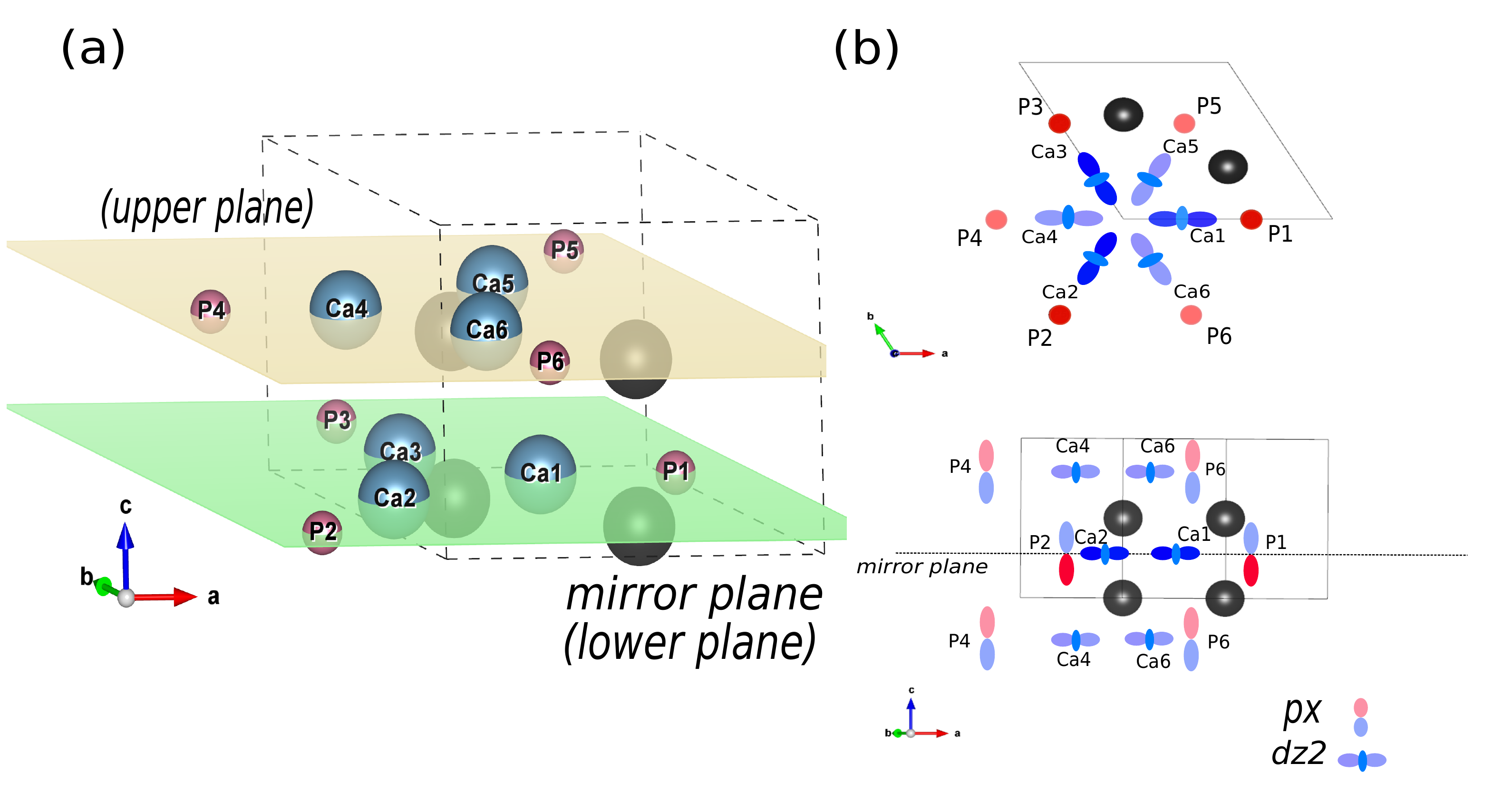}
\end{center}
\begin{center}
\includegraphics[clip,width=0.95\columnwidth]{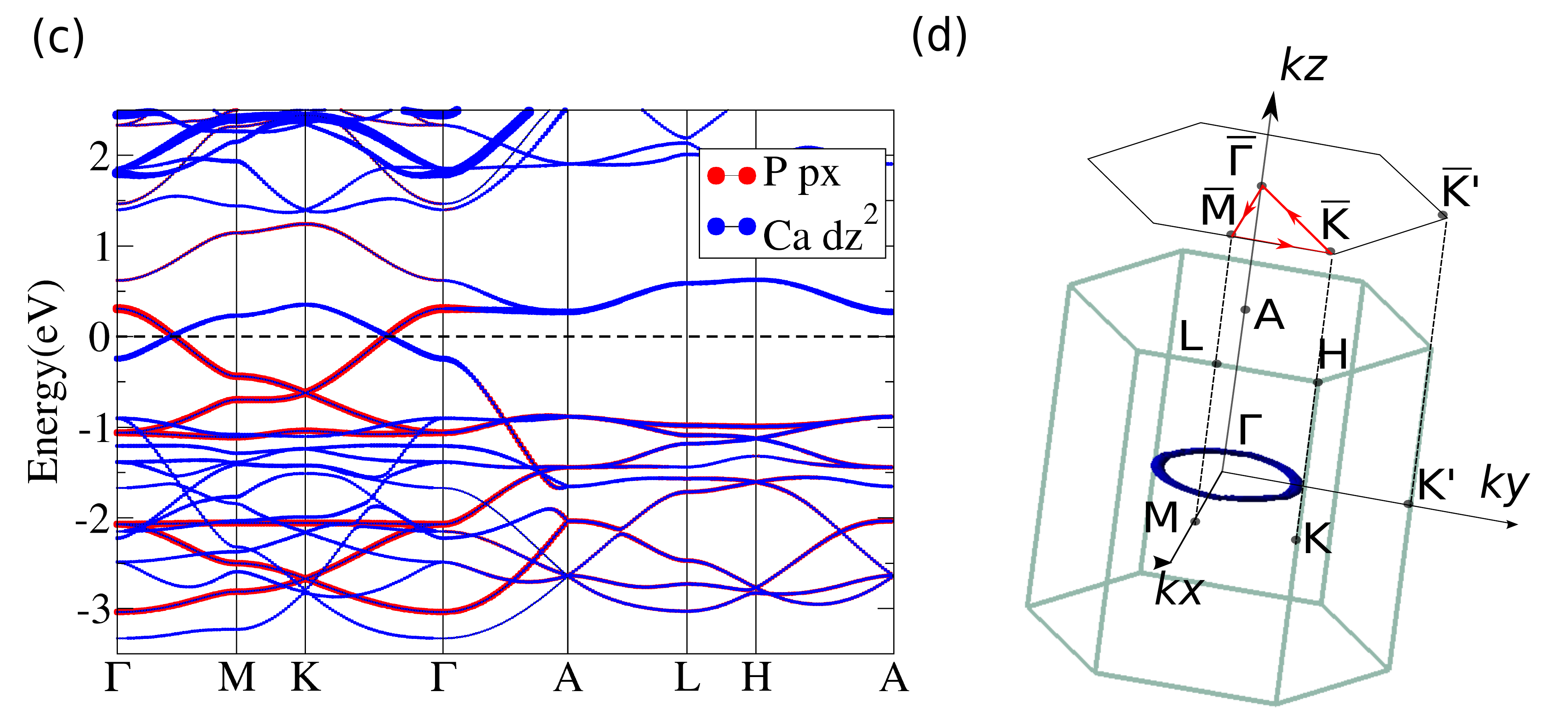}
\end{center}
  \caption{  \label{mm_Fig1}
Crystal structure and electronic bands of Ca$_3$P$_2$.
(a)~Crystal structure of Ca$_3$P$_2$, which contains two planes with three Ca atoms (blue) and three P atoms (red)
that are separated by interstitial Ca atoms (black).
The gray dashed lines indicate the unit cell. 
(b)~Top and side view of the crystal structure. The P-$p_x$ and \mbox{Ca-$d_{z^2}$} orbitals 
included in the tight-binding model are shown schemati\-cally.
% The dark blue (Ca) and the red (P) orbitals belong to atoms in the lower layer while the light blue and the pink orbitals lie in the upper layer.  
(c) Calculated electronic band structure of Ca$_3$P$_2$. The weights of the
P-$p_x$ and \mbox{Ca-$d_{z^2}$} orbitals that are located within the layers
are indicated by the width of the corresponding band. 
The weight of the \mbox{Ca-$d_{z^2}$} orbital 
is multiplied by two  to make it 
more visible on the scale of the plot.
(d) Fermi ring of Ca$_3$P$_2$ as obtained from the tight-binding model, Eq.~\eqref{tight_binding_ham}.
The bulk and surface Brillouin zones are outlined
by the green and black lines, respectively. 
}   
\label{FermiRing} 
\end{figure}

In this paper, by considering Ca$_3$P$_2$ as a representative example of a topological  semi-metal, we discuss the stability of topological Fermi lines in terms  
of crystalline topological invariants %e.g., mirror Chern or mirror winding numbers,
 that take on  nonzero quantized values. These topological numbers measure the global phase
structure of the electronic wavefuncitons in the presence of symmetry constraints.
We derive and compute the $\mathbb{Z}$- and $\mathbb{Z}_2$-type crystalline invariants for both a tight-binding model (Sec.~\ref{sec_tight-binding}) and a low-energy effective description of Ca$_3$P$_2$ (Sec.~\ref{sec_low_energy_theory}). It follows from our analysis that the four-fold degenerate Dirac ring of  Ca$_3$P$_2$ [Fig.~\ref{FermiRing}(d)] is protected 
against gap opening by reflection symmetry and SU(2) spin-rotation symmetry.
%(class A with $R$ of Ref.~\onlinecite{ChiuSchnyder14}).
The Dirac ring can be split into two two-fold degenerate Weyl rings 
by spin-rotation symmetry breaking perturbations, see Figs.~\ref{mm_Fig5} and~\ref{mm_Fig6}. We find that the
stability of both the Dirac ring and the Weyl ring are guaranteed by a $\mathbbm{Z}$-type mirror invariant (Sec.~\ref{sec_fermi_ring_protection}).
The Fermi ring of Ca$_3$P$_2$ can also be stabilized by
time-reversal symmetry combined with inversion,
instead of reflection,
in which case the  protection is due to a  $\mathbbm{Z}_2$-type topological number.  

Unlike in crystalline topological insulators~\cite{Teo:2008fk,Fu11,Chiu_reflection,Morimoto2013,Sato_Crystalline_PRB14}, the crystalline invariants for nodal line semi-metals are not directly linked with the appearance of surface states.
Nevertheless,  as we show in Sec.~\ref{sec_surface_states} and Fig.~\ref{SurfaceStatesCa3P2},
there appear topological ingap states at the surface of Ca$_3$P$_2$,
which arise from a quantized Berry phase, rather then the crystalline invariant.
Since the Berry phase is equal to $\pi$ for any closed path that interlinks with the Fermi line,
surface states with a 
nearly flat dispersion occur 
within two-dimensional regions of the surface Brillouin zone.
These surface states take the form of a drumhead  that is bounded by the projected Fermi lines (Fig.~\ref{SurfaceStatesCa3P2}).
We derive in Sec.~\ref{sec_relation_Berry_MZ} an important relation between the $\mathbbm{Z}$-type mirror invariant 
and the Berry phase, which establishes a direct
connection between the appearance of the nearly flat surface states and the
topological stability of the bulk Fermi line. It follows from this relation that
drumhead boundary states are a generic feature of topological nodal line semi-metals, 
occurring in both Weyl and Dirac ring systems (Figs.~\ref{SurfaceStatesCa3P2}, \ref{mm_Fig5}, and \ref{mm_Fig6}).

In the presence of disorder or interactions the  surface states of nodal line semi-metals can scatter and interact with quasiparticles in the bulk, since there is no full gap in the system. 
Hence, impurity scattering or electron-electron correlations might potentially destroy the
boundary modes. For  nearly flat surface states the effects of interactions are particularly strong,
since their large density of states enhances correlation effects.  
Hence, even relatively weak interactions may lead to exotic symmetry broken states at the surface, such as surface magnetism or surface superconductivity (Sec.~\ref{sec_summary_and_conclusions}).
Regarding the effects of disorder, we find that bulk impurities
do not destroy the surface states
as long as: (i) the disorder strength is considerably smaller than the energy gap separating
valence from conduction bands and (ii)
 the disorder respects  reflection symmetry on average.

The remainder of this paper is organized as follows. In Sec.~\ref{sec_tight-binding} we discuss
the topological features of nodal line semimetals in terms of a tight-binding model.
We start in Sec.~\ref{sec_derive_tight-binding} by
deriving a twelve band tight-binding Hamilltonian for Ca$_3$P$_2$ using maximally localized Wannier 
functions.  This is followed by a discussion of
the topological stability of the Dirac ring in Sec.~\ref{sec_fermi_ring_protection}.
%The surface state spectrum of Ca$_3$P$_2$ is presented in Sec.~\ref{sec_surface_states},
We show in Sec.~\ref{sec_surface_states} that
a non-zero quantized Berry phase
leads to the appearance of nearly flat surface states.
The relation
between the Berry phase and the crystalline topological invariant
is derived in Sec.~\ref{sec_relation_Berry_MZ}. 
Sec.~\ref{sec_TR_breaking} is devoted to the  study of time-reversal and inversion
breaking perturbations, which split the Dirac ring into two Weyl rings.
To show that the topological features discussed in Sec.~\ref{sec_tight-binding}
are generic to any nodal line semi-metal, we discuss in
Sec.~\ref{sec_low_energy_theory} an effective continuum model that describes
the low-energy physics near a general topological Fermi line.
We evaluate the crystalline invariant for this continuum model in Sec.~\ref{sec_connt_top_invariants2}.
%The $\mathbbm{Z}_2$ number that stabilizes the Fermi ring in the presence of
%time-reversal and inversion symmetry is presented in Sec.~\ref{sec_continuum_surface_states}.
In Sec.~\ref{sec_continuu_symm_breaking_sec} we study how time-reversal and inversion breaking terms split the Fermi line.
Finally, in Sec.~\ref{sec_summary_and_conclusions} we conclude the paper and give an outlook on future research.
Sec.~\ref{sec_summary_and_conclusions} also contains a brief discussion of the effects of disorder %and interactions 
on the topological surface states. Some technical details have been relegated to four appendices.

%%%%%%%%%%%%%%%%%%
%%%%%%%%%%%%%%%%%%%

%%%%%%%%%%%%%%%%%%%%%%%%
%%%%%%%%%%%%%%%%%%%%%%%%%

\section{Tight-binding calculations}
\label{sec_tight-binding}

In this section, we examine the band structure topology
of Ca$_3$P$_2$ in terms of a tight-binding model with twelve bands.
Although the analysis below is performed specifically for Ca$_3$P$_2$,  the principles discussed
in this section are valid more generally and can be applied to any material
with the same symmetries as Ca$_3$P$_2$.

%%%%%%%%%%%%%%%%%%%%%%%%%%%%%%%%
%%%%%%%%%%%%%%%%%%%%%%%%%%%%%%%%

\subsection{Tight-binding model for C$\text{a}_3$P$_2$}
\label{sec_derive_tight-binding}

 Recently, a new polymorph of Ca$_3$P$_2$ has been synthesized
which crystallizes in a hexagonal lattice structure with space group $P6_3/mcm$~\cite{xie_schoop_arXiv}.
Figures~\ref{FermiRing}(a) and \ref{FermiRing}(b)  display the crystal structure of this polymorph of  Ca$_3$P$_2$,
which contains two layers  with three Ca and three P atoms  separated
by four  interstitial Ca atoms. X-ray diffraction measurements show that the
Ca site  is only partially occupied, yielding a Ca$^{2+}$--P$^{3-}$ charge-balanced compound.

To determine the electronic band structure   
we perform first principles calculations with the WIEN2k code~\cite{blahaWien2k}
using as an input the experimental crystal structure of Ref.~\cite{xie_schoop_arXiv}.  
For the exchange-correlation functional we choose the 
generalized-gradient approximation of Perdew-Burke-Ernzerhof type~\cite{pbe_PRL_96}.
The full Brillouin zone is sampled by $21 \times 21 \times 22$ $k$-points
and the plane-wave  cut-off is set to $RK_{max}=7$.
We treat the partial occupancy of the Ca atoms  within the virtual crystal approximation~\cite{nordheim_VCA}.
Figure~\ref{FermiRing}(c) shows the calculated band
structure of Ca$_3$P$_2$ within an energy range of $\pm 3$~eV around the Fermi energy $E_\textrm{F}$.
To obtain the orbital character of the bands we introduce  a  local coordinate system
for each Ca and P site, whose definition is illustrated in Fig.~\ref{FermiRing}(b).
In each coordinate frame the $x$ axis is oriented along the $c$ direction, whereas
the $z$ axis lies with the $ab$ plane, pointing towards the lower left edge of the unit cell [Fig.~\ref{FermiRing}(b)].
With these definitions, we find that 
 the bands close to the Fermi energy mainly originate from the Ca-$d_{z^2}$ and  P-$p_x$  orbitals
 that are located within the layers [Fig.~\ref{FermiRing}(c)].
 The other orbitals of the in-plane atoms (Ca-$d_{xy}$, Ca-$d_{xz}$, Ca-$d_{yz}$, 
 Ca-$d_{x^2-y^2}$, P-$p_y$, and  P-$p_z$), as well as all the orbitals
 of the Ca interstitials, contribute insignificantly to the low-energy bands and can
 be neglected for the construction of the tight-binding model. 
% The basis wave functions are chosen as follows

 Guided by these observations, we use
the six Ca-$d_{z^2}$ and the six P-$p_x$ 
orbitals that are located within the two layers as a basis set
for the low-energy-tight binding model. 
Hence, the tight-binding  Hamiltonian is defined in terms of a twelve-component Bloch spinor
\begin{eqnarray} \label{Bloch_spinor}
\left| \psi_{\bf k}^{\alpha}  \right\rangle 
=
\frac{1}{\sqrt{N}} \sum_{\bf R} e^{ i {\bf k} \cdot ( {\bf R} + {\bf s}_\alpha ) }
\left| \phi^{\alpha}_{\bf R} \right\rangle ,
\end{eqnarray}
where $\alpha$ is the orbital index, ${\bf R}$ denotes  the lattice vectors,
and ${\bf s}_{\alpha}$ represents the position vectors of the six Ca ($\alpha = 1, \ldots, 6$)
and the six P sites ($\alpha = 7, \ldots 12$), as specified in Figs.~\ref{FermiRing}(a) and~\ref{FermiRing}(b). 
For completeness, the numerical values of the position vectors ${\bf s}_{\alpha}$ are given in Table~\ref{table_postion_vectors} of Appendix~\ref{appencix_A}.
At this stage of the discussion, we ignore the spin degree of freedom  of the Bloch spinor,
since spin-orbit coupling is negligibly small for the light elements Ca and P.
Using the  spinor~\eqref{Bloch_spinor}, we construct the  matrix elements of the Bloch Hamiltonian as
\begin{eqnarray} \label{eq:TBelements} \label{tight_binding_ham}
H^{\alpha \beta} ( {\bf k}  )
=
\langle \psi^{\alpha}_{\bf k} | H  | \psi^{\beta}_{\bf k} \rangle
=
\sum_{\bf R} e^{ i {\bf k} \cdot ( {\bf R} + {\bf s}_{\alpha} - {\bf s}_{\beta} )}   t^{\alpha  \beta}_{\bf R}  ,
\end{eqnarray}
%The matrix elements of Hamiltonian read
%\begin{eqnarray}
%\bra{\psi_{\bf k, \beta}}H\ket{\psi_{\bf k, \alpha}} =
%\sum_{\bf R,n} t_n e^{i\bf{k}(R+\tau_\alpha-\tau_\beta)},
%\label{eq:TBelements}
%\end{eqnarray}
where $t^{\alpha \beta}_{\bf R}$ is the hopping amplitude from orbital $\alpha$
in the unit cell at the origin  to orbital $\beta$ in the unit cell at position ${\bf R}$. 
To simplify the form of the matrix elements~\eqref{tight_binding_ham} and have a single-valued Hamiltonian, we absorb 
a momentum dependent phase factor in the definition of the basis orbitals, i.e., 
we let $| \psi^{\alpha}_{\bf k} \rangle \to   e^{i {\bf k} \cdot {\bf s}_{\alpha}} | \psi^{\alpha}_{\bf k} \rangle$.
We observe that Hamiltonian~\eqref{eq:TBelements}� has a nested block structure 
\begin{eqnarray} \label{def_ham_block_structure}
 H  ( {\bf k} )
=
\begin{pmatrix}
\textrm{H}_{\textrm{Ca} \textrm{Ca}} & \textrm{H}_{\textrm{Ca} \textrm{P}} \cr
\textrm{H}_{\textrm{P} \textrm{Ca}}  & \textrm{H}_{\textrm{P} \textrm{P}} \cr
\end{pmatrix},
\; 
H_{ij}  =
\begin{pmatrix}
h_{ij}^{\textrm{ll}} & h_{ij}^{\textrm{lu}}  \cr
h_{ij}^{\textrm{ul}} & h_{ij}^{\textrm{uu}} \cr  
\end{pmatrix},
\end{eqnarray}
where the sub-blocks $h_{ij}^{mn}$ with fixed  $i,j \in \left\{ \textrm{Ca}, \textrm{P} \right\}$ and  fixed $m,n \in \left\{ \textrm{l}, \textrm{u} \right\}$ are $3 \times 3$ matrices.
The outer blocks $\textrm{H}_{ij}$ represent hopping processes among and between the Ca and P orbitals, 
whereas the inner blocks ($h_{ij}^{\textrm{uu}}$, $h_{ij}^{\textrm{ll}}$) and
($h_{ij}^{\textrm{lu}}$, $h_{ij}^{\textrm{ul}}$) describe intralyer and interlayer hoppings, respectively.
The detailed form of the matrix elements $h_{ij}^{mn}$ is specified in Appendix~\ref{appencix_matrix_elements},
where we also describe how the hopping parameter values are determined from 
a maximally localized Wannier function (MLWF)  method~\cite{kuneWannier90,marzariRMP12}. 

\begin{figure}[t!]
 \begin{center} 
\includegraphics[clip,width=0.491\columnwidth]{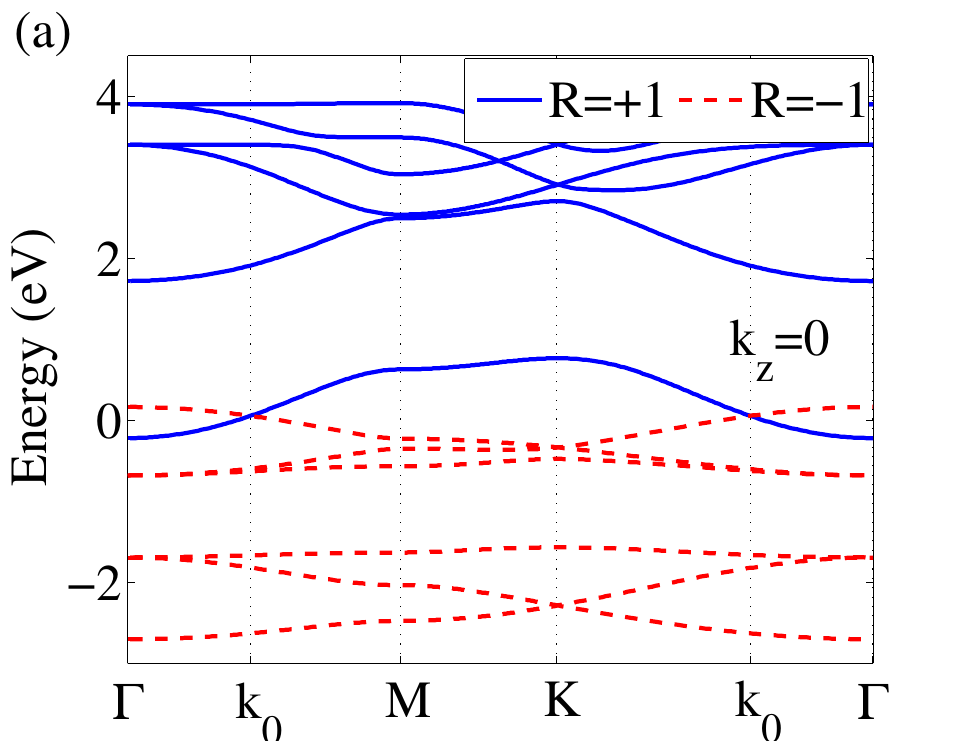}
\includegraphics[clip,width=0.491\columnwidth]{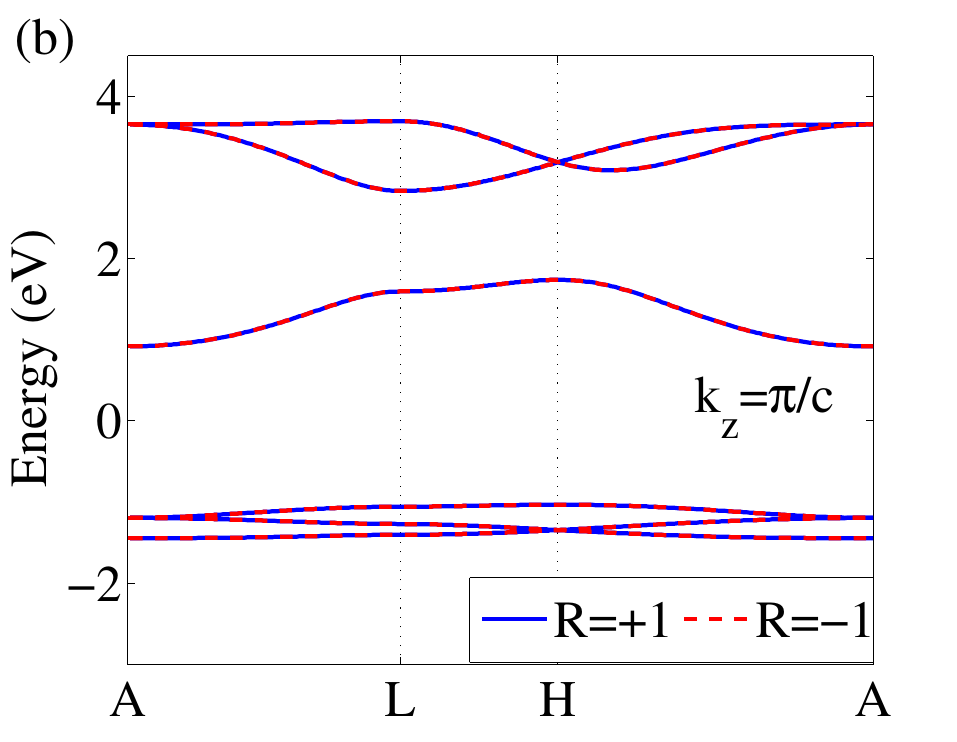}
 \end{center} 
 \caption{ \label{FigureZwei} \label{mm_Fig2}
Band structure of the tight-binding model. 
Panels (a) and (b)  show the energy bands of Hamiltonian~\eqref{tight_binding_ham}
along high-symmetry lines within the mirror planes $k_z=0$ and $k_z = \pi / c$, respectively
[cf.\ Fig.~\ref{FermiRing}(d)] .
The reflection eigenvalues of the bands are indicated by color, with
blue and red corresponding to   $R=+1$ and  $R=-1$, respectively.
}
\end{figure}

In Fig.~\ref{FermiRing}(d) we plot the energy isosurface of Hamiltonian~\eqref{tight_binding_ham} at $E=E_{\textrm{F}} \pm20$~meV, which shows that the tight-binding model correctly captures the fourfold degenerate Dirac ring of Ca$_3$P$_2$. 
Comparing the first-principles band structure of Fig.~\ref{FermiRing}(c) with the tight-binding bands
displayed in Fig.~\ref{FigureZwei}, we find that the tight-binding model
closely reproduces the bands with dominant Ca-$d_{z^2}$ and P-$p_x$  orbital character. 
In particular, the linear dispersion close to the Dirac ring agrees well with the
first-principles results.

\subsubsection{Symmetries}

%The Fermi-ring in Ca$_3$P$_2$ is protected by the reflection symmetry along the $z$ direction.

As we will see in the following sections, time-reversal, inversion, reflection,
and SU(2) spin-rotation symmetry play a crucial role for the
protection of the Dirac ring.
Let us therefore discuss how these symmetries act on the tight-binding Hamiltonian.

First of all, since we did not include the spin degree of freedom
in Eq.~\eqref{tight_binding_ham}, the tight-binding model is fully SU(2) spin-rotation invariant.
That is, our model is
diagonal in spin space with Hamiltonian~\eqref{tight_binding_ham} 
representing the diagonal element. 
As a consequence, the time-reversal operator is simply given by
the identity matrix times the complex conjugation operator $\mathcal{K}$, i.e., 
$T= \mathbbm{1} \mathcal{K}$, which
acts on the Hamiltonian as
\begin{eqnarray} \label{def_TRsym}
T^{-1} H(-\bk) T  =H(\bk) .
\end{eqnarray}
Hence, Hamiltonian~\eqref{tight_binding_ham} belongs to symmetry class AI, since $T^2 = +1$.
According to the classification of Ref.~\onlinecite{chiu_review15}  Fermi rings in this symmetry class are unstable
in the absence of  lattice symmetries. 
However, as we will discuss below, reflection symmetry or a combination of inversion
with time-reversal symmetry can produce a topological protection of the Dirac ring.

The two layers of the crystal structure of Ca$_3$P$_2$, indicated in green and brown in Fig.~\ref{FermiRing}(a),  are reflection planes. For brevity, we only discuss the lower reflection plane [colored in green in Fig.~\ref{FermiRing}(a)], but
the following analysis also holds, mutatis mutandis, for the upper plane.
The invariance of the tight-binding Hamiltonian~\eqref{tight_binding_ham} under reflection
about the lower plane implies
\begin{subequations} \label{def_reflection_sym}
\begin{eqnarray}
R^{-1}(k_z)H(k_x, k_y, -k_z) R(k_z)=H(k_x,k_y, k_z) ,
\qquad  \label{reflection_eq}
\end{eqnarray}
with the $k_z$-dependent reflection operator 
\begin{eqnarray}
R ( k_z )
&=&
\tau_z
\otimes
e^{i\frac{k_z}{2} (\rho_z- \rho_0) c}
\otimes \mathbbm{1}_{3 \times 3} 
\nonumber\\
&=&
\tau_z \otimes 
\begin{pmatrix}
1 & 0 \cr
0 & e^{+ i k_z c } \cr
\end{pmatrix}
 \otimes \mathbbm{1}_{3 \times 3}   ,
\end{eqnarray}
\end{subequations}
where $c$ is the length of the lattice vector 
along the (001) direction.
Here, the two sets of Pauli matrices $\tau_\alpha$ and $\rho_\alpha$
describe the orbital (Ca-$d_{z^2}$, P-$p_x$) and the layer ($\textrm{l}$, $\textrm{u}$) degrees of freedom, respectively.
The form of the reflection operator $R (k_z)$ follows from the observations that (i)
the P-$p_x$ orbitals are odd under reflection, while the Ca-$d_{z^2}$ orbitals are even;
and (ii) the mirror symmetry maps the orbitals in the upper layer to the next unit cell,
which gives rise to the phase factor $e^{+ i k_z c}$.
% the shift of one unit cell for orbitals in the upper layer after the mirror operation.
Finally, we find that the tight-binding model is also inversion symmetric.
That is, Hamiltonian~\eqref{tight_binding_ham}  satisfies 
\begin{eqnarray}  \label{eq_Inversion_eq}
I^{-1} H(-\bk)  I  =H(\bk),
\end{eqnarray}
with the  spatial inversion operator $I =\tau_0 \otimes \rho_x \otimes \mathbbm{1}_{3 \times 3}$.

%%%%%%%%%%%%%%%%%%%%%%%%%%%%%%%%
%%%%%%%%%%%%%%%%%%%%%%%%%%%%%%%%

\subsection{Topological protection of the Fermi ring}
\label{sec_fermi_ring_protection}

%note that for $k_z$ in the mirror plane $R$ commutes with $T$, and hence
%belongs to class  AI with $R_+$ of Ref.~\onlinecite{ChiuSchnyder14})
%

%We first consider Ca$_3$P$_2$ preserves SU(2) symmetry. This spin $1/2$ can be simply decomposed to two identical spinless systems. We mainly focus on one of the spinless systems to show the Fermi ring is protected by reflection symmetry in the $z$ direction and lock the ring at $k_z=0$ with the reflection operator in {\cb section II}. 

%Furthermore, $R^2=\bI$ leads to that the eigenvalues of $R$ is $\pm 1$.

Let us now discuss how reflection symmetry~\eqref{def_reflection_sym} leads to the topological 
protection of the Dirac ring.  
First, we observe that for ${\bf k}$ within the reflection plane $k_z=0, \pi$ the mirror operator $R (k_z) $ commutes with 
Hamiltonian~\eqref{tight_binding_ham}, i.e., $\left[ R ( k_z), H (k_x, k_y, k_z) \right] = 0$ for $k_z = 0, \pi$. 
Therefore, it is possible to block-diagonalize $H ( {\bf k} )$ within the mirror planes with respect to $R$.
In this block-diagonal basis each eigenstate of $H ( {\bf k} )$ 
has either mirror eigenvalue $R=+1$ or $R=-1$.
As we can see from Fig.~\ref{mm_Fig2}(a), the two bands that cross
at the Dirac point have opposite mirror eigenvalues, 
which prevent hybridization between them.
In other words, any term that couples the two bands 
breaks reflection symmetry.
The stability of the band crossing is guaranteed by a mirror
invariant of type $M \mathbbm{Z}$~\cite{ChiuSchnyder14}.
This mirror index is given
by the difference of occupied states with eigenvalue $R=+1$
on either side of the Dirac ring, i.e., 
\begin{eqnarray} \label{eq_mirror_invariant}
N^0_{M \mathbb{Z} }
=
n^{+,0}_{\textrm{occ}} ( | {\bf k}_{\parallel}  | > k_0 )
- 
n^{+,0}_{\textrm{occ}} ( | {\bf k}_{\parallel}  | < k_0 ) , 
\end{eqnarray}
where ${\bf k}_{\parallel} = ( k_x, k_y) $  is the in-plane momentum
and
\begin{eqnarray} \label{eq_occ_number}
n^{+,0}_{\textrm{occ}} (   {\bf k}_{\parallel}   )
=
\left\{
\begin{array}{l l}
1, & \, | {\bf k}_{\parallel} |<  k_0 \; \; \textrm{(inside the ring)} \\
0, & \, | {\bf k}_{\parallel} | >  k_0 \; \; \textrm{(outside the ring)}
\end{array}
\right.
\end{eqnarray}
 denotes the number of occupied states
at $({\bf k}_{\parallel}, 0)$
in the mirror eigenspace $R=+1$.

In passing, we note that Hamiltonian~\eqref{tight_binding_ham} is  a member of symmetry 
 class  AI with $R_+$ in the terminology of Ref.~\onlinecite{ChiuSchnyder14},
 since $T^2=+1$ and $R$ commutes with $T$. However, nodal lines with codimension $p=2$ in class
 AI with $R_+$ are unstable, since for this class there does not exist any zero-dimensional invariant 
 defined at time-reversal invariant momenta within the mirror plane. Nevertheless, the Dirac band crossing is protected, since the Hamiltonian
 can also be viewed as
 a member of  class A with $R$. The mirror invariant for the latter
  class [i.e., Eq.~\eqref{eq_occ_number}], which is defined for any in-plane momentum ${\bf k}_{\parallel}$, can be non-zero even in the presence of time-reversal symmetry.
Besides reflection symmetry, the product of inversion
and time-reversal symmetry $IT$ also protects the Dirac line. This will be discussed at the end of Sec.~\ref{sec_surface_states}
and in Sec.~\ref{sec_continuum_mirror_invariants} in terms of a low-energy continuum model.

\subsection{Surface states and Berry phase} \label{sec_surface_states}

\begin{figure}[t!]
 \begin{center} 
\includegraphics[clip,width=0.98\columnwidth]{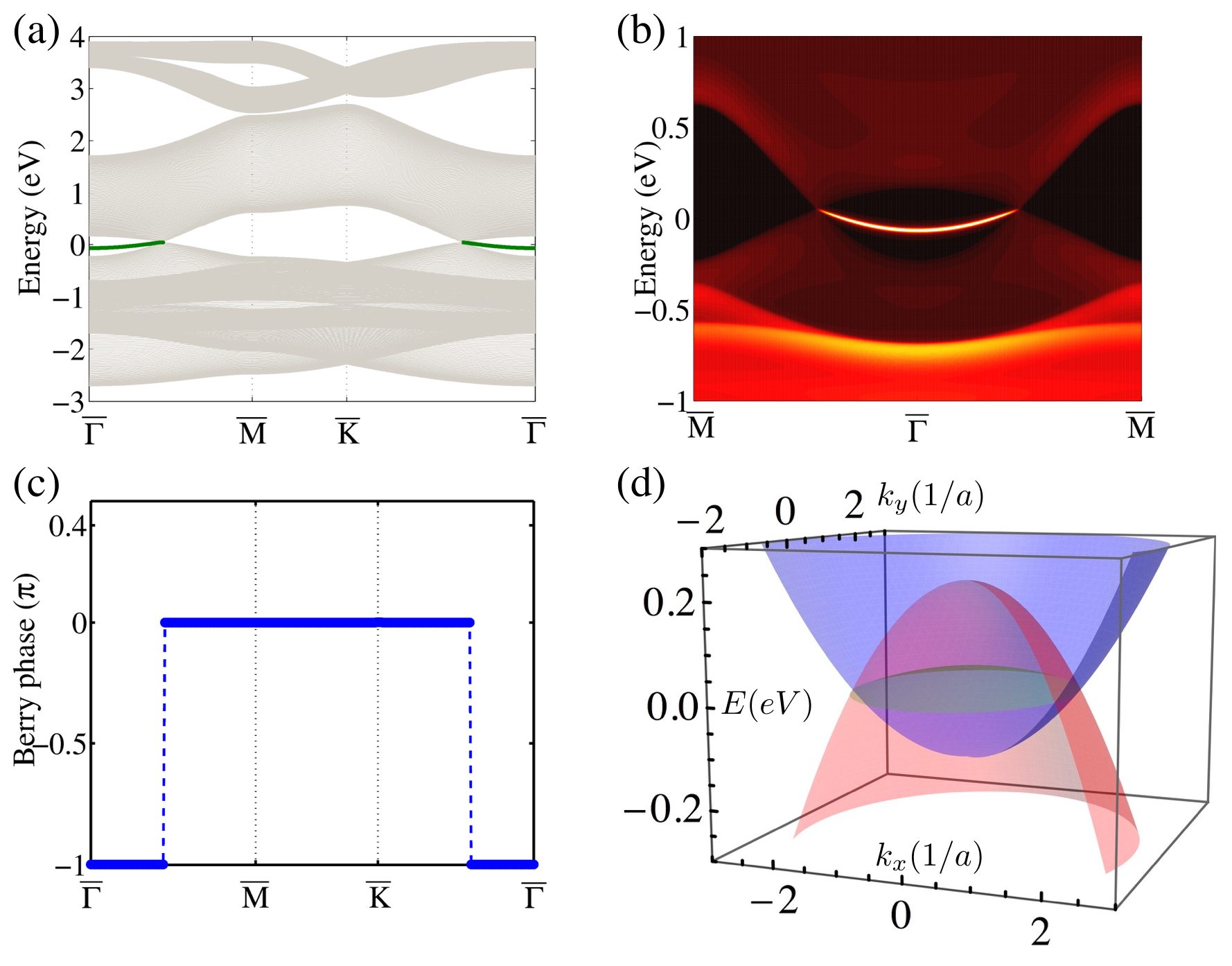}
 \end{center} 
 \caption{ \label{Figure_surface_states} \label{mm_Fig3}
Drumhead surface states and Berry phase. 
(a)~Surface band structure of Ca$_3$P$_2$  as obtained from the tight-binding model~\eqref{tight_binding_ham} for the (001) surface in slab geometry with 60 unit cells. The surface state is highlighted in green.
(b)~Momentum-resolved surface density of states of Hamiltonian~\eqref{tight_binding_ham}   for the (001) surface.
White and dark red correspond to high and low density, respectively.
(c)~Variation of the Berry phase~\eqref{eq_berry_phase}
of Hamiltonian~\eqref{tight_binding_ham}
along high-symmetry lines of the (001) surface Brillouin zone [see Fig.~\ref{FermiRing}(d)].
(d)~Surface spectrum of the low-energy effective model~\eqref{eq_low_energy_spinless} for the (001) face as a function of 
surface momenta $k_x$ and $k_y$.
The bulk states at $k_z=0$ with reflection eigenvalues $R=+1$ and $R=-1$ are colored
in blue and red, respectively.
The drumhead surface state is indicated by the green area. 
}
 \label{SurfaceStatesCa3P2}
\end{figure}

In this section, we present the 
surface spectrum of Ca$_3$P$_2$  as obtained from the tight-binding model~\eqref{tight_binding_ham}
and show that, due to a non-zero  Berry phase, there
appear nearly flat ingap states at the surface.
Figure~\ref{mm_Fig3}(a) displays the surface band structure 
for the (001) surface in a three-dimensional slab geometry with 60 unit cells.
The surface momentum is varied 
along a high-symmetry path, which is drawn in red in 
 the surface Brillouin zone of Fig.~\ref{mm_Fig1}(d).
Using an iterative Green's function method~\cite{sancho_iterative_Green}
we compute the momentum resolved surface density of states for a semi-infinite (001) slab, 
which is shown in Fig.~\ref{mm_Fig3}(b).
As indicated by  the green area in Fig.~\ref{mm_Fig3}(d) and by the green and yellow lines in Figs.~\ref{mm_Fig3}(a) and~\ref{mm_Fig3}(b), respectively, the surface state is nearly dispersionless,  taking the shape
of a drumhead that is bounded by the projected Dirac ring. 
We note that nearly or completely flat surface states have recently also been studied
in photonic crystals~\cite{Lu_photonic_NatPhoton_2015}, in noncentrosymmetric superconductors~\cite{SchnyderRyuFlat,tanaka_sato_nagaosa_review_12,schnyderPRL_13,SchnyderJPCM15}, in bernal graphite~\cite{volovik_bernal_graphite},
and in topological crystalline insulator heterostructures~\cite{tangFu_natPhys14}.

In contrast to crystalline topological insulators the surface states of the semimetal~\eqref{tight_binding_ham}
are not directly related to the mirror invariant~\eqref{eq_mirror_invariant}, but 
are connected to a non-zero  Berry phase.
To make this connection explicit, we decompose the (001) slab considered in Fig.~\ref{mm_Fig3} into a
family of one-dimensional systems parametrized by the in-plane momentum ${\bf k}_{\parallel} = (k_x, k_y)$.
For fixed ${\bf k}_{\parallel}$, the Berry phase is defined as
\bee \label{eq_berry_phase}
\mathcal{P} ( {\bf k}_{\parallel} ) 
=
-i \sum_{E_j< E_{\textrm{F}} }   \int_{-\pi}^{\pi}     \bra{u_{j} ( {\bf k} ) } \partial_{k_z} \ket{u_{j} ( {\bf k} )}  dk_z ,
\ee
where the sum is over  filled Bloch eigenstates $| u_j ( {\bf k} ) \rangle$ of Hamiltonian~\eqref{tight_binding_ham}.
As was shown by King-Smith and Vanderbilt~\cite{kingSmithPRB93b},  the
Berry phase $\mathcal{P} ({\bf k}_{\parallel})$  is related to the charge $q_{\textrm{end}}$ at the end of the one-dimensional system
with fixed in-plane momentum ${\bf k}_{\parallel}$, i.e., 
\begin{eqnarray} \label{eq_end_charge_berry}
q_{\textrm{end}} =  \frac{e}{ 2 \pi } \mathcal{P} ( {\bf k}_{\parallel} ) \; \; \textrm{mod} \; \; e . 
\end{eqnarray}
Hence, when $\mathcal{P} ( {\bf k}_{\parallel }) \ne 0$   an ingap  state appears
at ${\bf k}_{\parallel}$ in the surface Brillouin zone. 
For the tight-binding Hamiltonian~\eqref{tight_binding_ham} we find that there are 
two different symmetries which each quantize the  Berry phase~\eqref{eq_berry_phase}
to $0$ or $\pi$, namely, the reflection symmetry~\eqref{def_reflection_sym} and  
the product of time-reversal and inversion symmetry $IT$, see Appendix~\ref{appendixB}.
In Fig.~\ref{mm_Fig3}(c) we numerically compute $\mathcal{P} ( {\bf k}_{\parallel} )$
using the tight-binding wave functions of Hamiltonian~\eqref{tight_binding_ham}.
We obtain that the Berry phase equals $\pi$ for ${\bf k}_{\parallel}$ inside the projected Dirac ring, while
it is zero for ${\bf k}_{\parallel}$ outside the ring. 
This indicates that surface states occur within the projected Dirac ring, which is
in agreement with the surface spectrum of Figs.~\ref{mm_Fig3}(a) and~\ref{mm_Fig3}(b). 
The Berry phase is defined modulo $2 \pi$, since large gauge transformations
of the wave functions change it by $2 \pi$. As a result, $\mathcal{P}$
protects only single,
but not multiple, surface states at a given ${\bf k}_{\parallel}$.

Remarkably due to the  $IT$ symmetry, the Berry phase~$\mathcal{P}$
along any closed loop in the three-dimensional Brillouin zone is quantized (see Appendix~\ref{appendixB}).
This allows  us to interprete the Berry phase as a
topological invariant which guarantees the stability of the Dirac line in the presence of the  $IT$ symmetry.
That is, for a loop interlinking with the Dirac ring, we find that $\mathcal{P}= \pm \pi$
which shows that the Dirac band crossing is protected by the product of inversion with
time-reversal symmetry. The Berry phase represents a   $\mathbb{Z}_2$-type invariant,
since it is defined only up to multiples of $2 \pi$.
In contrast, the mirror number~\eqref{eq_mirror_invariant} 
is a $\mathbb{Z}$-type invariant, which can take on any integer number.
Therefore, only the mirror invariant~$N_{M \mathbb{Z} }$ can give rise to the stability of multiple Dirac lines
at the some location in the Brillouin zone.

%which ensures the stability of the band crossing against gap opening.
%Not gauge invariant under large gauge transformations.

%%%%%%%%%%%%%%%%%%%%%%%%%%%%%%%%
%%%%%%%%%%%%%%%%%%%%%%%%%%%%%%%%

\subsection{Relation between Berry phase \\ and mirror invariant}
\label{sec_relation_Berry_MZ}

%Although the Berry phase and $N_\bZ$ independently characterize the two different physical features, they are deeply related by the reflection symmetry operator

The analysis of the previous section suggests that the topological stability of the Dirac ring
is closely related  to the appearance of surface states.
In order to put this connection on a firmer footing, we present here a relation between the mirror
invariant and the Berry phase $\mathcal{P}({\bf k}_{\parallel})$. 
Namely, we find that 
\begin{subequations} \label{relation_MZ_berry}
\begin{eqnarray}
(-1)^{n^{+,0}_{\textrm{occ}} ( {\bf k}_{\parallel} ) + n^{+,\pi}_{\textrm{occ}}( {\bf k}_{\parallel} )} e^{i \partial R}
=
e^{i \mathcal{P} ( {\bf k}_{\parallel} ) } 
\end{eqnarray}
for all in-plane momenta ${\bf k}_{\parallel} = (k_x, k_y)$, where 
\begin{equation}
 \partial R 
=
i \sum_{E_j < E_{\textrm{F}}} 
\int_0^\pi
 \langle u_{j} ({\bf k} ) | 
R^{\dag} (k_z) \left[ \partial_{k_z} R (k_z)  \right]
 | u_{j} ({ \bf k} )  \rangle
d k_z
\end{equation}
\end{subequations}
denotes the change in phase of the reflection operator $R (k_z) $ along the reflection direction $k_z$.
The invariants $n^{+,0}_{\textrm{occ}} ( {\bf k}_{\parallel} )$ and $n^{+,\pi}_{\textrm{occ}}( {\bf k}_{\parallel} )$
correspond to the number of occupied states
at $({\bf k}_{\parallel}, 0)$ and $({\bf k}_{\parallel}, \pi)$, respectively,
with mirror eigenvalue $R=+1$. 
Formula~\eqref{relation_MZ_berry}, whose proof is derived in Appendix~\ref{appendixB}, is one of the main results of this paper.
For concreteness we have assumed in~\eqref{relation_MZ_berry} that reflection symmetry $R(k_z)$ maps $z$ to $-z$. But relationn~\eqref{relation_MZ_berry}  is valid more generally, i.e., for any reflection symmetric semimetal, in particular also for line-node materials with strong spin-orbit coupling, such as PbTaSe$_2$~\cite{cava_PbTaSe2_PRB_14,bian_hasan_arXiv_15}.
 
 We observe that in general the reflection operator only depends on the momentum along the reflection
 direction [i.e., on $k_z$ in the case of Eq.~\eqref{def_reflection_sym}], but is independent of the in-plane momenta ${\bf k}_{\parallel}$.
Hence, we infer from Eq.~\eqref{relation_MZ_berry} that when the mirror invariant $n^{+,0}_{\textrm{occ}} ( {\bf k}_{\parallel} )$ 
[or $n^{+,\pi}_{\textrm{occ}} ( {\bf k}_{\parallel} )$] changes by one as the in-plane momentum ${\bf k}_{\parallel}$ is moved across the topological Dirac line, 
the Berry phase increases by $\pi$, since $\partial R$ does not depend on ${\bf k}_{\parallel}$. 
As a consequence,  a drumhead surface state appears
either inside or outside the projected Dirac ring. 
This proofs the direct connection between the stability of the Dirac ring and the existence of  drumhead surface states.
For the tight-binding model of Ca$_3$P$_2$, Eq.~\eqref{tight_binding_ham}, we find that
the phase change $\partial R$ of the reflection operator~\eqref{def_reflection_sym} evaluates 
to $3 \pi$ independent of ${\bf k}_{\parallel}$. Figure~\ref{mm_Fig2}(b) shows that 
the number of occupied states with momentum $({\bf k}_{\parallel}, \pi)$ and
mirror eigenvalue $R=+1$ is $n^{+,\pi}_{\textrm{occ}} (   {\bf k}_{\parallel}   ) = 3$
for all ${\bf k}_{\parallel}$.
Using relation~\eqref{relation_MZ_berry} together with Eq.~\eqref{eq_occ_number}, it follows that 
 the Berry phase $\mathcal{P}$ equals $\pi$ inside and 0 outside the Dirac ring, which agrees
with the explicit calculation of $\mathcal{P}$, see Fig.~\ref{mm_Fig3}(c).

In closing this section, we note that for certain highly symmetric lattice models~\cite{Hughes:2011uq,Chiu_reflection}
the reflection operator $R$
is completely momentum independent, in which case formula~\eqref{relation_MZ_berry} simplifies to
\begin{eqnarray}
\left[ n^{+,0}_{\textrm{occ}} ( {\bf k}_{\parallel} ) + n^{+,\pi}_{\textrm{occ}}( {\bf k}_{\parallel} ) \right] \pi
=
 \mathcal{P} ( {\bf k}_{\parallel} )   \ (\rm{mod}\ 2\pi),
\end{eqnarray}
for all ${\bf k}_{\parallel}$~\cite{footnote1}. Hence, in this case the Berry phase, and therefore the location of the surface states, is fully determined by the mirror invariant~\eqref{eq_occ_number}.
This is useful, since the mirror number~\eqref{eq_occ_number} is easier to compute than the Berry phase, for
which one needs to determine the momentum dependence of the tight-binding wave functions.

%%%%%%%%%%%%%%%%%%%%%%%%%%%%%%%%
\begin{figure}[t!]
 \begin{center} 
 \includegraphics[clip,width=0.98\columnwidth]{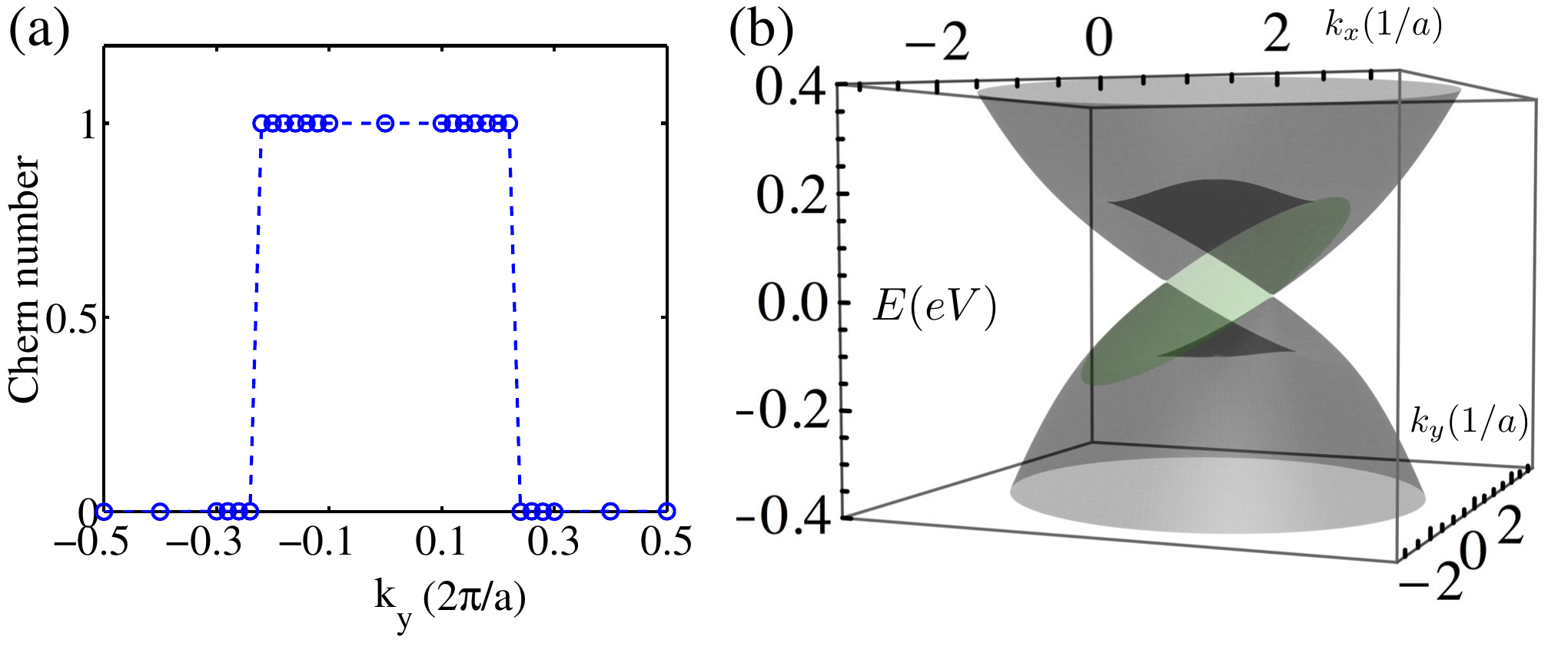}
 \end{center} 
 \caption{ \label{Figure_surface_states} \label{mm_Fig4}
Arc surface state and spin Chern number.
(a)~\mbox{$k_y$ dependence} of the
spin Chern number~\eqref{spin_chern_number} of Hamiltonian~\eqref{tight_binding_ham} in the
presence of the mirror and time-reversal symmetry breaking perturbation~\eqref{eq_mirror_breaking_perturb}.
(b)~Surface and bulk spectra of the low-energy model~\eqref{eq_low_energy_spinless} perturbed by  
the mass term~\eqref{eq_cont_TR_R_broken} % {\cb ($d(k_x-k_y)\tau_x/\sqrt{2}$)  as
with $d=0.9~\rm{eV}\AA$ and $\theta_0 = - \pi /4$, which breaks reflection and time-reversal symmetry.
The bulk states and the arc state at the (001) surface are indicated in  gray and green, respectively.
}
 \label{Weyl_points_fig}
\end{figure}
%%%%%%%%%%%%%%%%%%%%%%%%%%%%%%%%

\subsection{Symmetry-breaking perturbations}
\label{sec_TR_breaking}

We  have seen that the stability of the Dirac ring of Ca$_3$P$_2$
is protected by SU(2) spin-rotation symmetry, reflection symmetry, and the product of inversion and time-reversal symmetry $IT$.
In this section, we study how the breaking of these symmetries 
modifies the bulk and surface spectrum of Ca$_3$P$_2$.

\subsubsection{Reflection and time-reversal symmetry breaking}

First, we consider a reflection and time-reversal breaking perturbation with the 
following nonzero matrix elements
\begin{subequations} \label{eq_mirror_breaking_perturb}
\begin{eqnarray}
\bra{\psi^{1}_{\bk}} H \ket{\psi^{9}_{\bk }} = + 0.2\sin( {\bf k}\cdot {\bf r}_0)
\end{eqnarray}
and 
\begin{eqnarray}
\bra{\psi^4_{\bk}} H \ket{\psi^{12}_{\bk}} = -0.2\sin( {\bf k} \cdot {\bf r}_0),
\end{eqnarray}
\end{subequations}
where ${\bf r}_0 =(0.5, 0.5, 0)$ is a vector within the reflection plane along the diagonal direction. 
This term is odd in momentum ${\bf k}$ and couples the $d_{z^2}$ orbitals at the Ca1 and Ca4 sites  with the $p_x$ 
orbitals at the P3 and P6 sites [cf.~Figs.~\ref{mm_Fig1}(a) and~\ref{mm_Fig1}(b)]. 
It follows from Eqs.~\eqref{def_reflection_sym} and~\eqref{eq_Inversion_eq} that 
perturbation~\eqref{eq_mirror_breaking_perturb} breaks
reflection and time-reversal symmetry, but respects inversion symmetry. 
Therefore, Eq.~\eqref{eq_mirror_breaking_perturb} gaps out the Dirac ring except for two points along the 
diagonal direction  $(1,\,-1,\,0)$, where it vanishes [see Fig.~\ref{mm_Fig4}(b)].
These two gap closing points are Dirac nodes (or Weyl nodes, if one disregards the spin degree of freedom), whose stability is guaranteed by the spin 
Chern number~\cite{fukui_hatsugai_JPSJ_05}
\begin{eqnarray} \label{spin_chern_number}
C_{\textrm{s}} (k_y) 
=
\frac{1}{2 \pi i } \sum_{E_j < E_{\textrm{F}} } \int_{T^2} 
\left[ \partial_{k_x} A^{(j)}_z - \partial_{k_z} A^{(j)}_x \right] d k_x d k_z ,
\qquad
\end{eqnarray}
where $A^{(j)}_\mu =     \bra{u_{j} } \partial_{k_\mu} \ket{u_{j} }$ is the Berry connection.
We find that $C_{\textrm{s}} (k_y) $ evaluates to $+1$ for $k_x k_z$ planes 
inbetween the two Dirac points, while it is zero otherwise [Fig.~\ref{mm_Fig4}(a)].
By the bulk-boundary correspondence, the nonzero spin Chern number~\eqref{spin_chern_number} implies the appearance
of an arc  state in the surface Brillouin zone
connecting the projections of the two Dirac nodes [green area in Fig.~\ref{mm_Fig4}(b)].
As perturbation~\eqref{eq_mirror_breaking_perturb} is turned to zero, the arc  state transforms into the drumhead surface
state of Fig.~\ref{mm_Fig3}.

%The nodal line can be viewed as a continuum of Dirac nodes.

\subsubsection{Spin-rotation symmetry breaking}
\label{sec_tigh_bind_spin_sym_break}

Second, we study the effects of SU(2) spin-rotation symmetry breaking induced, for example,
by spin-orbit coupling. For Ca$_3$P$_2$ the spin-orbit interactions are negli\-gible due to the 
small atomic number of Ca and P. 
However, there are a number of topological semimetals with heavy elements, such as PbTaSe$_2$
and TlTaSe$_2$, for which spin-orbit coupling is strong.
Spin-orbit interactions can modify the energy spectrum of
nodal line semimetals in two different ways: either they open up a
full gap in the spectrum, or they split the Dirac ring into two Weyl rings.
Here, we  study the latter possibility. In order to do so, we need
to explicitly include the spin degree of freedom in Hamiltonian~\eqref{def_ham_block_structure}, i.e., we
consider
\begin{eqnarray} \label{tight_binding_soc}
\hat{H} ( {\bf k} ) =
H ( {\bf k} ) \otimes  \sigma_0 + \hat{H}_{\textrm{sb}} ( {\bf k} ) ,
\end{eqnarray}
where  $\sigma_0$ operates in spin space and $\hat{H}_{\textrm{sb}}   $ represents
a spin-rotation symmetry breaking term, which we specify below.
Time-reversal symmetry acts on $\hat{H} $ according to Eq.~\eqref{def_TRsym}, but with the modified 
time-reversal operator 
%$\hat{T} =  i \mathbbm{1} \otimes \sigma_y \mathcal{K}$. 
$\hat{T} = T  \otimes i \sigma_y$.
Similarly, the reflection operator and the spatial inversion operator are changed
to $\hat{R} = R \otimes \sigma_z$ and $\hat{I} = I \otimes \sigma_0$, respectively.
To split the four-fold degenerate Dirac ring of Eq.~\eqref{tight_binding_soc} into two two-fold degenerate Weyl rings,
it is necessary to also break  time-reversal or inversion symmetry, besides  spin-rotation
symmetry.

\begin{figure}[tb]
\begin{center}
\includegraphics[clip,width=0.98\columnwidth]{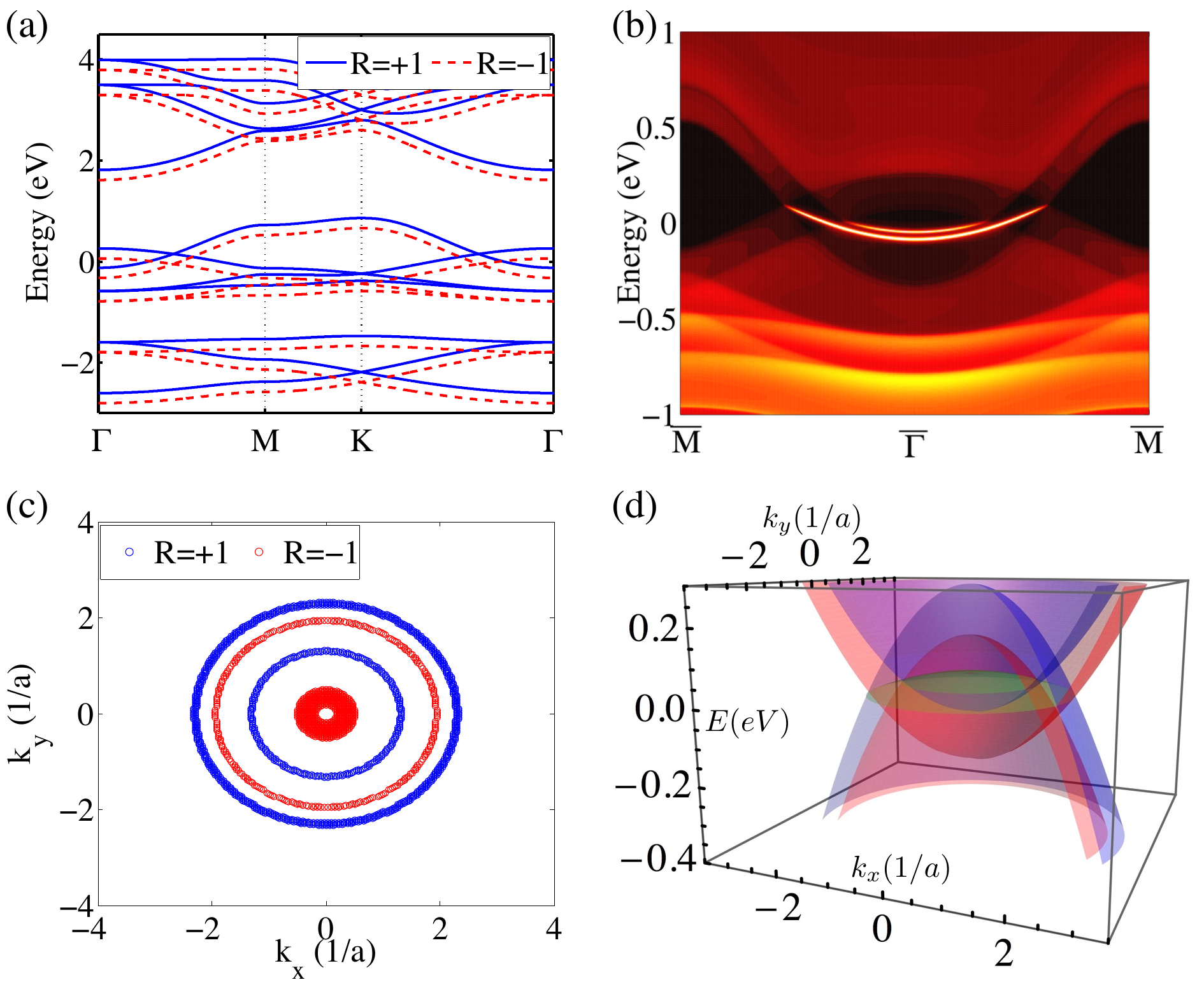}
\end{center}
  \caption{ \label{fig_TRS_breaking}  \label{mm_Fig5}
Bulk bands and drumhead surface states of a spinful time-reversal breaking
line-node semimetal.
Panels (a) and (b) show the
bulk bands and the surface density of states of Hamiltonian~\eqref{tight_binding_soc} in the
presence of the staggered Zeeman term~\eqref{eq_staggered_zeeman}  with $h_z = 0.1$~\textrm{eV}.
The momentum in panel (a) is varied within the mirror plane $k_z=0$
 along high-symmetry lines of the Brillouin zone.
(c) Energy isosurfaces of Hamiltonian~\eqref{tight_binding_soc}  with $h_z = 0.1$~eV at $E_F \pm5 $~meV and $k_z=0$. 
(d)~Surface and bulk spectra of the low-energy effective model~\eqref{Hspinful} 
perturbed by the time-reversal breaking term~\eqref{TRS-breaking} with $\nu_\parallel h^z_{\textrm{eff}}=0.07~\rm{eV}$. 
The drumhead states at the (001) surface are colored in green.
The reflection eigenvalues of the bulk bands at $k_z=0$ in panels (a), (c), and (d) are indicated by color, with blue and red corresponding to 
$R = +1$ and $R = -1$, respectively. }
\end{figure}

\paragraph{Time-reversal breaking perturbation}
The staggered Zeeman field
\begin{eqnarray} \label{eq_staggered_zeeman}
\hat{H}_{\textrm{sb}} ( {\bf k} )  = h_z \, \tau_z \otimes \rho_0 \otimes \mathbbm{1}_{3 \times 3} \otimes   \sigma_z
\end{eqnarray}
%with $h_z=0.1$.
breaks both time-reversal and spin-rotation symmetry,
but satisfies inversion and reflection symmetry. 
It describes an external staggered magnetic field with opposite
signs on the Ca and P sites.
According to the terminology of Ref.~\cite{ChiuSchnyder14}, Hamitlonain~\eqref{tight_binding_soc} perturbed by Eq.~\eqref{eq_staggered_zeeman} is a member of
class A with $R$, which exhibits an integer number of equivalence classes distinguished
by a mirror invariant.
In Figs.~\ref{mm_Fig5}(a) and~\ref{mm_Fig5}(c)
we present the bulk energy bands of Hamiltonian~\eqref{tight_binding_soc} with an applied staggered Zeeman field
of strength $h_z =0.1$~eV.
The bulk spectrum displays two Weyl rings, whose stability is guaranteed by the mirror number~\eqref{eq_mirror_invariant}.
Figures~\ref{mm_Fig5}(b) and~\ref{mm_Fig5}(d) show the surface energy spectrum at the (001) face.
We find that there are two drumhead surface states which are bounded
by the projections of the two Weyl rings.
%dispersing across the $\bar{\Gamma}$ point of the surface Brillouin zone. 
In accordance with the discussion of Secs.~\ref{sec_surface_states} and~\ref{sec_relation_Berry_MZ} [cf.~Eq.~\eqref{relation_MZ_berry}]
the single surface state that appears
between the projections of the outer and inner Weyl rings is protected by the Berry phase~\eqref{eq_berry_phase}, which takes on the nonzero quantized value $\mathcal{P} = \pm \pi$.  The two surface states that exist inside the projection of the inner Weyl ring, on the other hand, 
are topologically unstable.

%%%%%%%%%%%%%%%%%%%%%%%%%%%%%%%%
%%%%%%%%%%%%%%%%%%%%%%%%%%%%%%%%

\begin{figure}[tb]
\begin{center}
\includegraphics[clip,width=0.98\columnwidth]{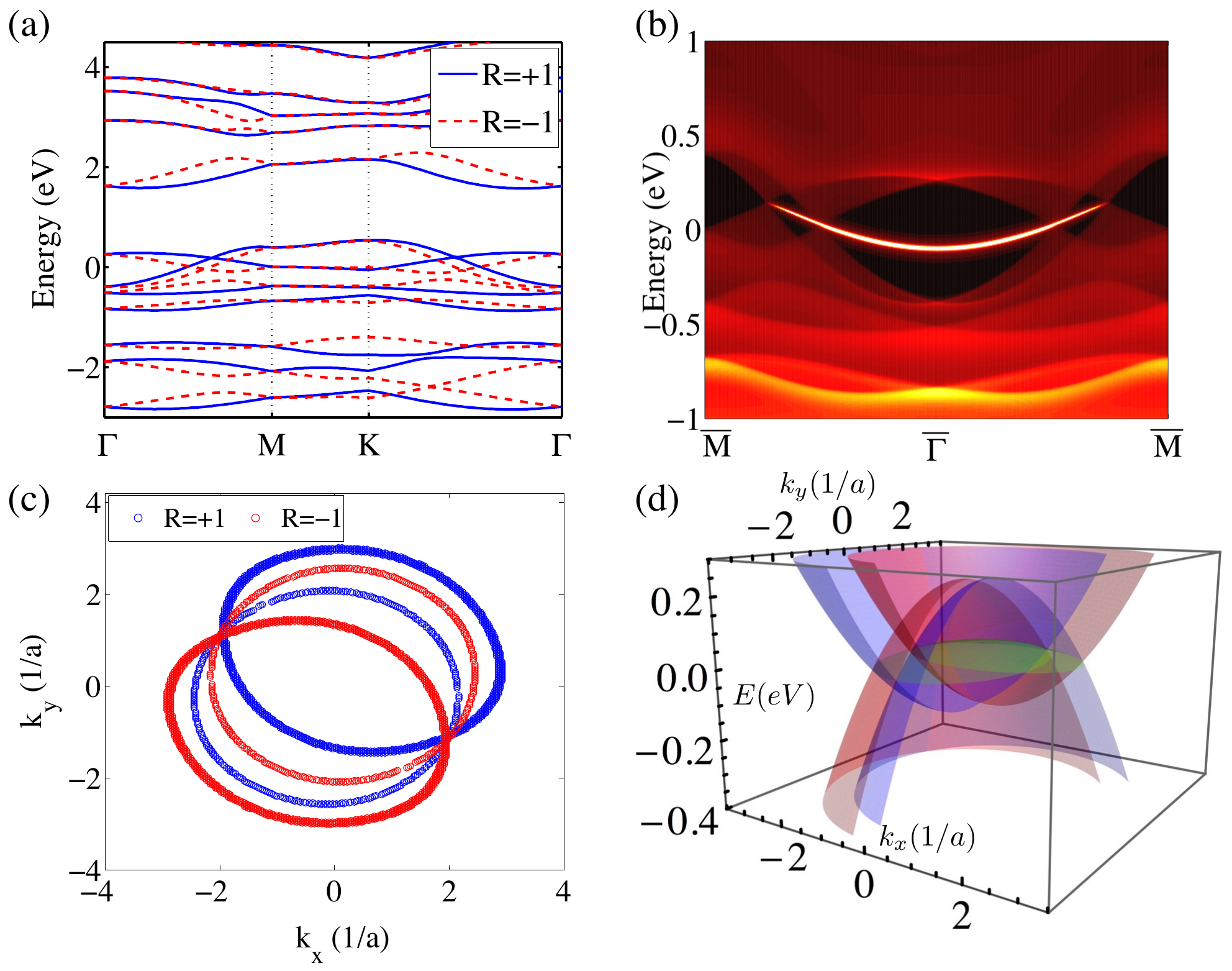}
\end{center}
  \caption{ \label{fig_inversion_sbreaking}  \label{mm_Fig6}
Bulk bands and drumhead surface states of 
a spinful inversion breaking line-node semimetal.
Panels (a) and (b) display the bulk bands and the surface density of states of
tight-binding model~\eqref{tight_binding_soc} in the presence of the inversion breaking term~\eqref{eq_Inversion_breaking_perturb}.
The momentum in panel (a) is varied within the mirror plane $k_z=0$ along high-symmetry lines.
(c)  Energy isosurfaces of Hamiltonian~\eqref{tight_binding_soc} perturbed by Eq.~\eqref{eq_Inversion_breaking_perturb} at $E_F \pm5$~meV and $k_z=0$.
(d) Surface and bulk spectra of the low-energy effective model (\ref{Hspinful}) 
perturbed by the inversion breaking term~\eqref{inversion-breaking} with $\delta=0.025~\rm{eV}\AA$.
The drumhead states at the (001) surface are indicated in green.
The mirror eigenvalues of the bulk bands at $k_z=0$ in panels (a), (c), and (d) are represented by color, with blue
and red corresponding to $R=+1$ and $R=-1$, respectively.
 }
\end{figure}

\paragraph{Inversion breaking perturbation}
To break inversion and spin-rotation symmetry we consider a 
perturbation with the following nonzero matrix elements 
\begin{subequations} \label{eq_Inversion_breaking_perturb}
\begin{equation}
\bra{\psi^{1}_{{\bf k} \sigma}} \hat{H} \ket{\psi^{6}_{{\bf k}  \sigma}} 
=
 + 0.6i \sgn (\sigma ) e^{ i{\bf k}\cdot  ( {\bf s}_6 - {\bf s}_1 ) }
\left[ 1 + e^{i {\bf k} \cdot \hat{\bf e}_z } \right] \qquad \quad
\end{equation}
and 
\begin{equation}
\bra{\psi^7_{{\bf k} \sigma   }   } \hat{H} \ket{\psi^{12}_{{ \bf k} \sigma }} 
= 
-0.3i \sgn ( \sigma) e^{ i{\bf k} \cdot ( {\bf s}_{12} - {\bf s}_7 + {\bf R}_{110} ) } \left[ 1 + e^{i {\bf k} \cdot \hat{\bf e}_z } \right] , \qquad \quad
\end{equation}
\end{subequations}
% ${\bf r}_{1\pm} =(-0.2029, -0.2029, \pm 0.5)$, ${\bf r}_{2 \pm } =(0.378, 0.378, \pm 0.5)$.
where $| \psi^{\alpha}_{{\bf k} \sigma} \rangle$ denotes the Bloch spinor with orbital index $\alpha$ and spin index $\sigma = \pm$.
The vectors ${\bf s}_{\alpha}$ are the position vectors of the atoms in the unit cell
and are given  in Table~\ref{table_postion_vectors} of Appendix~\ref{appencix_A}. 
Perturbation~\eqref{eq_Inversion_breaking_perturb} couples the orbitals at the Ca1 and  P1 sites with
the orbitals at the Ca6 and P6 sites, respectively.
Using Eqs.~\eqref{def_TRsym}, \eqref{def_reflection_sym}, and \eqref{eq_Inversion_eq} 
one can check that the term~\eqref{eq_Inversion_breaking_perturb} satisfies reflection and time-reversal symmetry, but breaks inversion 
symmetry.
Since $\hat{T}^2=-1$ and  $ \{ \hat{T} , \hat{R}  \} =0$, Hamiltonian~\eqref{tight_binding_soc} perturbed
by Eq.~\eqref{eq_Inversion_breaking_perturb} is a member of 
class AII with $R_-$ of Ref.~\cite{ChiuSchnyder14}, for which a
 mirror invariant can be defined.
The bulk bands at $k_z=0$ of Hamiltonian~\eqref{tight_binding_soc} in the presence 
of the inversion-breaking term~\eqref{eq_Inversion_breaking_perturb}
are presented in Figs.~\ref{mm_Fig6}(a) and~\ref{mm_Fig6}(b). 
We observe that the Dirac ring is split into two Weyl rings, 
which intersect on the $( \sqrt{3}, -1, 0)$ axis.
As in the previous cases, the Weyl nodal lines
are protected by the nonzero mirror number~\eqref{eq_mirror_invariant}.  
Figures~\ref{mm_Fig6}(b) and~\ref{mm_Fig6}(d) show 
the surface spectrum at the (001) surface, which exhibits
two drumhead surface states.
As before, we find that only the single surface state
which occurs between the projections of the inner and outer rings 
is protected by the Berry phase~\eqref{eq_berry_phase}.

%Again, surface states emanates from crossing of bands with different reflection eigenvalues.

%%%%%%%%%%%%%%%%%%%%%%%%%%%%%%%%
%%%%%%%%%%%%%%%%%%%%%%%%%%%%%%%%

%%%%%%%%%%%%%%%%%%
%%%%%%%%%%%%%%%%%

%%%%%%%%%%%%%%%%%%%
%%%%%%%%%%%%%%%%%%%

\section{Low-energy continuum theory of nodal line semimetals}
\label{sec_low_energy_theory}

In this section we present a low-energy effective theory for a general topological nodal line semimetal
with  time-reversal, reflection, and inversion symmetry.
The form of this low-energy description is universal, since
it is entirely dictated by symmetry. We start by discussing Dirac rings, which arise in semi\-metals
with conserved SU(2) spin-rotation symmetry. Spin rotation breaking semimetals with Weyl nodal lines
will be discussed in Sec.~\ref{sec_continuum_ham_with_SOC}.

Consider the following low-energy Hamiltonian with spin-rotation symmetry
\bee \label{eq_low_energy_spinless}
H_{\rm{eff}} ({\bk})= \nu_\parallel (k_{\parallel}^2-k_0^2)\tau_z+\nu_z k_z\tau_y + f({\bf k}) \tau_0  ,
\ee 
which describes a Dirac ring within the $k_z=0$ plane, located at $k_\parallel^2:= k^2_x + k^2_y=k_0^2$.
In Eq.~\eqref{eq_low_energy_spinless} we suppress the spin degree of freedom, since any spin-dependent terms
are  forbidden by symmetry.
The Pauli matrices $\tau_i$ operate in orbital space
and the function $f ( {\bf k})$ is restricted by symmetry to be even in ${\bf k}$.
We assume that  $f({\bf k})= \nu_0 (k_\parallel^2 -k_0^2)+V_0$, neglecting any 
terms of higher order in ${\bf k}$.
To make a connection with the previous section,
we fit the parameters $\nu_0$, $\nu_{\parallel}$, $\nu_z$, $k_0$, and $V_0$  
to the low-energy band structure of the DFT calculations of 
Sec.~\ref{sec_derive_tight-binding} [see Fig.~\ref{mm_Fig1}(c)].
We find that the momentum parameter $k_0$ equals $k_0= 0.206$~\AA$^{-1}$, 
the chemical potential is $V_0=0.095$~eV, and
the velocities are given by $\nu_0=-0.993$~eV\AA$^2$,
$\nu_\parallel =4.34$~eV\AA$^2$, and $\nu_z= 2.50$~eV\AA.  
Employing Eqs.~\eqref{def_TRsym}, \eqref{def_reflection_sym}, and \eqref{eq_Inversion_eq}, 
one can show that the low-energy Hamiltonian $H_{\textrm{eff}}$ satisfies 
time-reversal, reflection, and inversion symmetry, with
the symmetry operators
  $T_{\textrm{eff}} =\tau_0 \mathcal{K}$, $R_{\textrm{eff}} =\tau_z$, and $I_{\textrm{eff}} =\tau_z$, respectively.
Before we discuss in the next section the topological stability of the Dirac line~\eqref{eq_low_energy_spinless}, let us remark that $H_{\textrm{eff}}( {\bf k} )$ can be converted in a straightforward manner to 
a lattice model, see Appendix~\ref{toy lattice}.
In Figs.~\ref{mm_Fig3}(d), \ref{mm_Fig4}(b), \ref{mm_Fig5}(d), and \ref{mm_Fig6}(d)  we use the lattice version of Eq.~\eqref{eq_low_energy_spinless} 
to plot the surface states. Observe that there are some minor differences
in the shape of the surface states between the thigh-binding model~\eqref{tight_binding_ham}
and the effective theory~\eqref{eq_low_energy_spinless} [compare Fig.~\ref{mm_Fig3}(b) with Fig.~\ref{mm_Fig3}(d)]. We attribute
this difference to the omission of longer range hopping terms in Eq.~\eqref{eq_low_energy_spinless}.

%may be used for future studies of the effects of
%disorder and interactions on the surface states of $H_{\textrm{eff}}( {\bf k} )$.

%which however we leave for future work.

%which mainly captures physics as the energy is near the nodal lines and 
%To exhibit the stability of  the nodal lines, we study the  $4\times 4$ low energy Hamiltonian reduced from the effective tight-binding Hamiltonian {\cb (which equation?)} of Ca$_3$P$_2$ with $1/2$-spin
 
\subsection{Topological protection of the Fermi ring}
\label{sec_connt_top_invariants2}

As mentioned in Sec.~\ref{sec_fermi_ring_protection},  Dirac nodal lines
are protected by either reflection symmetry $R$ or the product 
of inversion with time-reversal symmetry $IT$.
Let us now discuss this in terms of the low-energy theory~\eqref{eq_low_energy_spinless}.

\setcounter{paragraph}{0}
\paragraph{$\mathbb{Z}$ classification due to reflection symmetry}

Considering only reflection symmetry and disregarding the spin degree of freedom, Hamiltonian~\eqref{eq_low_energy_spinless} belongs to class A
with $R$.
Since the codimension of the Dirac ring is $p=2$, it  
 is classified by an $M \mathbb{Z}$ invariant (see Table~II of Ref.~\cite{ChiuSchnyder14}), i.e., by the
 mirror number~\eqref{eq_mirror_invariant}, which measures the difference of occupied states with mirror eigenvalue $R_{\textrm{eff}} = +1$ on either side of the Dirac ring. The
two bands that cross at the nodal line have opposite reflection eigenvalues,
which prohibits hybridization between them. 
 Indeed, we find that the hybridization term $\tau_x$ breaks reflection symmetry $R_{\textrm{eff}}$.
 We note that the mirror invariant~\eqref{eq_mirror_invariant} is of $\mathbb{Z}$ type and can therefore
 protect multiple Dirac crossings in the Brillouin zone.
 To verify this for the low-energy model~\eqref{eq_low_energy_spinless}, 
 we enlarge the matrix dimension of Hamiltonian $H_{\textrm{eff}}$ by considering
$H_{\textrm{eff}}\otimes \mathbbm{1}_{n\times n}$, which respects reflection
symmetry with the enlarged reflection operator $R^{\prime}_{\textrm{eff}} = R_{\textrm{eff}}  \otimes \mathbbm{1}_{n \times n}$.
Hybridization terms for the enlarged Hamiltonian are of the form $\tau_x\otimes A$,
with $A$ an arbitrary  $n \times n$  Hermitian matrix. However, since  
$(R^{\prime}_{\textrm{eff}} )^{-1} ( \tau_x \otimes A ) R^{\prime}_{\textrm{eff}} = - \tau_x \otimes A$, all of these terms
 break reflection symmetry $R^{\prime}_{\textrm{eff}}$.
 
%Reflection symmetry $R^{\prime}_{\textrm{eff}}$ forbids any hybridization terms for the enlarged Hamiltonian,
%which are of the form 
%Furthermore, the nodal line is locked in $k_z=0$ since $\sigma_y$, which is the only term moving the rings out of $k_z=0$ plane, is forbidden by the reflection symmetry. 

\paragraph{$\mathbb{Z}_2$ classification due to $IT$ symmetry}

Besides reflection, also the product of inversion with time-reversal symmetry $I_{\textrm{eff}}T_{\textrm{eff}}$ prohibits hybridization between 
the two bands, since the hybridization term $\tau_x$ is not invariant under $I_{\textrm{eff}}T_{\textrm{eff}} = \tau_z \mathcal{K}$. 
But in the presence of $I_{\textrm{eff}} T_{\textrm{eff}}$, 
Dirac nodal lines are classified as $\mathbb{Z}_2$ instead of $M \mathbb{Z}$.
To see this, consider two copies of Hamiltonian $H_{\textrm{eff}}$, i.e.,
$H_{\textrm{eff}} \otimes \mu_0$, 
which satisfies $IT$ symmetry with the doubled operator
 $I_{\textrm{eff}} T_{\textrm{eff}} \otimes  \mu_0$.
Here, 
$\mu_{\alpha}$ denotes an additional set of Pauli matrices.
The Dirac rings of this doubled Hamiltonian are topologically unstable, since
the symmetry-preserving term $\tau_x\otimes \mu_y$ gaps out the nodal lines.  
As discussed at the end of Sec.~\ref{sec_surface_states},
the product of inversion with time-reversal symmetry $IT$ quantizes the Berry phase $\mathcal{P}$
to $0$ or $\pi$~\cite{Nodal_line_response_Taylor,FangFu_PRB15}. 
Hence, $\mathcal{P}$ can be interpreted as a $\mathbbm{Z}_2$ topological invariant
that guarantees the stability of the nodal ring. In contrast to the mirror invariant, the 
integration path that enters in the definition of this $\mathbbm{Z}_2$ number [cf.~Eq.~\eqref{eq_berry_phase}],
is not confined to the mirror plane. For any integration path that interlinks with the nodal line, $\mathcal{P} = \pm \pi$ 
signals the stability of the Dirac ring.

%	\emph{IT} symmetry also quantizes the Berry phase to $0$ or $\pi$ The integration path of the Berry path now is not limited only in the reflection symmetric path. Along any closed loop path, the Berry phase is still quantized. When we choose a path enclosing the nodal line, $\pi$ phase  leads to the protection of the nodal line. }	
%Furthermore, {\cb the polarization} can be extended to any possible directions. This is the reason that surface modes are coming from the projection of bulk nodal lines in any surfaces, which is in agreement with the following discussion of Weyl semimetals. 

In closing we note that, while the low-energy theory~\eqref{eq_low_energy_spinless} accurately captures the topological stability of 
the nodal ring of a given semi-metal, it does not necessarily correctly reproduce the location of the drumhead surface state.
That is, in order to determine whether the drumhead surface state is located  inside or outside the projected Dirac ring, 
it is necessary to compute the Berry phase of all the occupied states. This information is
not contained in the low-energy model~\eqref{eq_low_energy_spinless}, cf.~Appendix~\ref{toy lattice}.

 %The low energy theory mainly shows the symmetry protection of the nodal lines. The location of surface modes, however, cannot be predicted from the low energy theory since $\pi$ Berry phase, that leads to the presence of surface modes, can be only obtained by knowing the other occupied  states. 
%	

\subsection{Symmetry-breaking perturbations}
\label{sec_continuu_symm_breaking_sec}

In analogy to the discussion of Sec.~\ref{sec_TR_breaking}, we now study how different symmetry breaking
perturbations transform the Dirac ring~\eqref{eq_low_energy_spinless} into Dirac points
or Weyl rings.
	
\subsubsection{Reflection and time-reversal symmetry breaking}
\label{sec_continuum_surface_states}
\label{sec_continuum_mirror_invariants}

The Dirac line node of $H_{\textrm{eff}}$ can be deformed into two Dirac points by
the perturbation
\bee	 \label{eq_cont_TR_R_broken}
	d \, \sin(\theta_\parallel - \theta_0) k_\parallel  \tau_x, 
\ee
which breaks reflection and time-reversal symmetry, but respects inversion symmetry.
Here, $\theta_\parallel=\tan^{-1}(k_y/k_x)$ and $k_{\parallel} = \sqrt{ k_x^2 + k_y^2}$
denote polar angle and  absolute value of the in-plane momentum ${\bf k}_{\parallel}$, respectively.
The term~\eqref{eq_cont_TR_R_broken}Ê gaps the Dirac ring except at  
${\bf k} = \pm k_0 ( \cos \theta_0,  \sin \theta_0, 0)$. These two 
gap closing points are 
Dirac nodes with opposite chiralities, which are protected by the spin Chern number~\eqref{spin_chern_number}.
Due to the bulk-boundary correspondence an arc state  appears at the surface,  connecting
the  projected Dirac points  % at ${\bf k}_{\parallel} = \pm k_0 ( \cos \theta_0, \sin \theta_0) $
 in the surface Brillouin zone.
This is illustrated in Fig.~\ref{mm_Fig4}(b), where we set  $\theta_0 = - \pi / 4$ and $d=0.9$~eV\AA ,
which mimics the effects of perturbation~\eqref{eq_mirror_breaking_perturb}
for the tight-binding Hamiltonian~\eqref{tight_binding_ham}.

% are continuously connected

From the arc surface state of the above Dirac semimetal one can infer 
the existence of the drumhead surface state of $H_{\textrm{eff}}$,
since the two transform into each other by letting $d$ tend to zero in Eq.~\eqref{eq_cont_TR_R_broken}.
Moreover, the one-dimensional set of Dirac nodes, induced by Eq.\eqref{eq_cont_TR_R_broken} and parametrize by $\theta_0$, can be
interpreted as the Dirac ring of $H_{\textrm{eff}}$.
That is, as we let $\theta_0$ in Eq.~\eqref{eq_cont_TR_R_broken} run from $0$ to $\pi$ a nodal ring is created.
For each fixed $\theta_0$ there is an arc surface state connecting
 the two points ${\bf k}_{\parallel} = \pm k_0 ( \cos \theta_0,   \sin \theta_0 )$ in the surface Brillouin zone.
 Hence, a drumhead surface state is generated when $\theta_0$ is varied from $0$ to $\pi$.
From this argument one infers that  drumhead states also appear at surfaces for which the
Berry phase~\eqref{eq_berry_phase} is not quantized (cf.~Sec.~\ref{sec_surface_states}), since the appearance of  arc states does not depend on any crystal symmetries.

%We note that this argument does not guarantee the location of the surface states, which can be in the nodal line or outside the nodal line. 

%However, in reality $SU(2)$ symmetry, which is fragile, can be easily broken by the presence of spin orbit coupling and magnetic field. 

\subsubsection{Spin-rotation symmetry breaking}	
%\subsubsection{Dirac nodal line is decomposed to Weyl nodal lines by symmetry breaking}
\label{sum_breaking_terms_continuum}
\label{sec_continuum_ham_with_SOC}

In the absence of SU(2) spin-rotation symmetry, the Dirac ring of $H_{\textrm{eff}}$ is topologically unstable.
To discuss this, we consider as in Sec.~\ref{sec_tigh_bind_spin_sym_break} a spinful version of Hamiltonian~\eqref{eq_low_energy_spinless}
\begin{eqnarray} \label{Hspinful}
\hat{H}_{\textrm{eff}} ( {\bf k} ) 
=
\hat{H}_{\textrm{eff}} ( {\bf k} )  \otimes \sigma_0
+
\hat{H}^{\textrm{sb}}_{\textrm{eff}} ( {\bf k} ) ,
%H_{\text{spinful}}({\bf k})=\nu_\parallel(k_{\parallel}^2-k_0^2)\tau_z\sigma_0+\nu_z k_z\tau_y\sigma_0 + f({\bf k})\bm{1}  , 
\end{eqnarray}
where the Pauli matrices $\sigma_\alpha$ describe the spin degree of freedom
and $\hat{H}^{\textrm{sb}}_{\textrm{eff}}   $ denotes a spin-rotation symmetry breaking term.
$\hat{H}_{\textrm{eff}}$ is invariant under the same symmetries as the 
spinful tight-binding Hamiltonian~\eqref{tight_binding_soc}. That is, it satisfies time-reversal, reflection, and inversion 
symmetry with the operators $\hat{T}=\tau_0 \otimes i \sigma_y \mathcal{K}$, 
$\hat{R}=\tau_z \otimes \sigma_z $, and $\hat{I}=\tau_z \otimes \sigma_0$,
respectively.
%chiral symmetry is broken by $f(\bf k)\bm{1}$. 
%We find that  $SU(2)$ the total $N_{M\bZ}=1$ everywhere so the reflection protection is absent; 
We find that, the Dirac nodal lines of $\hat{H}_{\textrm{eff}}$ can be gapped 
out by the spin-rotation symmetry breaking mass terms $\tau_x \otimes \sigma_x$ 
and $\tau_x \otimes \sigma_y$, which preserve reflection symmetry $\hat{R}$
as well as   $\hat{I} \hat{T}$ symmetry.
These perturbations turn Hamiltonian~\eqref{Hspinful} into a trivial insulator. 
However, there exist also other spin-rotation symmetry breaking terms that deform the Dirac ring into two Weyl rings. 
These perturbation terms break either time-reversal symmetry or inversion symmetry.

%The nodal line is not locked at the mirror plane. We can simply add $\delta \sigma_y$, which preserves \emph{IT} symmetry but  breaks reflection symmetry, to shift the nodal line off the mirror plane.

\paragraph{Time-reversal breaking perturbation}
First, we add a spin-rotation and time-reversal breaking term to the Hamiltonian $\hat{H}_{\textrm{eff}}$,
which takes the form of a staggered Zeeman field
\bee \label{TRS-breaking}
\hat{H}^{\textrm{sb}}_{\textrm{eff}} ( {\bf k} )= \nu_\parallel h_{\textrm{eff}}^z   \tau_z  \otimes \sigma_z  .
\ee
This perturbation respects reflection and inversion symmetry. It splits the Dirac ring into two Weyl rings that are located within
the mirror plane $k_z=0$ at $ k_\parallel = \sqrt{k_0^2 \pm h^z_{\textrm{eff}} }$. 
The stability of these Weyl nodal lines is guaranteed by the mirror invariant~\eqref{eq_mirror_invariant},
which evaluates to 
\begin{eqnarray} \label{phaseFacGraph}
n^{+,0}_{\textrm{occ} } (k_\parallel)
=
  \left\{  \begin{array}{l l}
    1 ,& \,  k_\parallel < \sqrt{k_0^2- h^z_{\textrm{eff}} }   \\   
            0 ,&   \,  \sqrt{k_0^2- h^z_{\textrm{eff}} } < k_\parallel < \sqrt{k_0^2+ h^z_{\textrm{eff}} }   \\
            1 ,& \,   \sqrt{k_0^2+ h^z_{\textrm{eff}} } <  k_\parallel   
  \end{array} \right. . \quad
\end{eqnarray}
In Fig.~\ref{mm_Fig5}(d) we plot the surface spectrum of $H_{\textrm{eff}}$ in the presence
of the staggered Zeeman term with $\nu_\parallel h_{\textrm{eff}}^z = 0.07$~eV. There appear two drumhead surface states
which are bounded by the two projected Weyl rings.

%To simplify the problem we assume $f({\bf k})=0$ so the nodal lines are at the same Fermi level. The Weyl nodal lines indicate the boundary of the different Berry phases, which is integrated along the $z-$direction. The low energy theory, however, cannot lead to the exact value of the Berry phase, which between the two Weyl lines is $\pi$. The main goal to compute  the Berry phase is to know the location of the surface states. To obtain the Berry phase, we have to go back to the tight-binding model {\cb cite equation} since the occupied bands may make contributions for the Berry phase. The Berry phase can be either  computed directly in the tight-binding or obtained by counting reflection number $N_{M\bZ}$ at the reflection planes. We note $\partial R$ vanishes in spin-degenerate systems so $N_{M\bZ}$ can lead to the Berry phase by \eqref{invariant spin eqn}. 

\paragraph{Inversion breaking perturbation}

%%%%%%%%%%%%%%%%%%%%%%%%%
%%%%%%%%%%%%%%%%%%%%%%%%%

Alternatively, the Dirac ring can be split into Wely rings by an inversion breaking perturbation. 
To show this, we consider
\bee \label{inversion-breaking}
\hat{H}^{\textrm{sb}}_{\textrm{eff}} ( {\bf k} ) =	\delta (k_x + \sqrt{3} k_y)\tau_z \otimes \siz  ,
\ee
which respects reflection and  time-reversal symmetry.
In the presence of this term Hamiltonian~\eqref{Hspinful} exhibits two Weyl rings
within the mirror plane $k_z=0$ 
 with in-plane momenta given by the equation
 $(k_x\pm \delta/2)^2+(k_y\pm\sqrt{3}\delta/2)^2= k_0^2 +\delta^2$. 
 These two Weyl rings intersect on the $( \sqrt{3}, -1, 0)$ axis, where
 the gap term~\eqref{inversion-breaking} vanishes [cf.~Fig.~\ref{mm_Fig6}(c)].
We find again that these Weyl rings are protected by the 
mirror number~\eqref{eq_mirror_invariant}, with
\begin{eqnarray} \label{phaseFacGraph}
\begin{small}
n^{+,0}_{\textrm{occ}} (k_\parallel)
=
  \left\{  \begin{array}{l l}
    1 ,&   (k_x\pm \frac{\delta}{2})^2+ (k_y\pm \frac{\sqrt{3}\delta}{2})^2 > k_0^2 +\delta^2\ \& \\   
   &  (k_x\mp \frac{\delta}{2})^2+ (k_y\mp \frac{\sqrt{3}\delta}{2})^2 < k_0^2 +\delta^2 \\  
    0 ,&    \rm{otherwise}  
  \end{array} \right.  . \quad
  \end{small}
\end{eqnarray}
Fig.~\ref{mm_Fig6}(d) shows the surface spectrum of $\hat{H}_{\textrm{eff}}$ perturbed by Eq.~\eqref{inversion-breaking}.
As for the tight-binding model with the inversion-breaking term~\eqref{eq_Inversion_breaking_perturb}, there appear two drumhead surface states.
We note that PbTaSe$_2$~\cite{cava_PbTaSe2_PRB_14,bian_hasan_arXiv_15}
and TlTaSe$_2$~\cite{bian_hasan_arXiv_15b}  are examples of inversion breaking
semi\-metals with Weyl nodal lines. The low-energy physics of these materials can
be described by the effective theory~\eqref{Hspinful} perturbed by a term of the form~\eqref{inversion-breaking}.

%%%%%%%%%%%%%%%%%
%%%%%%%%%%%%%%%%%

%%%%%%%%%%%%%%%%
%%%%%%%%%%%%%%%%%

\section{Summary and discussion}
\label{sec_summary_and_conclusions}

In this paper we have studied the topological stability of Dirac and Weyl line nodes of three-dimensional
semimetals in the presence of reflection symmetry, time-reversal symmetry, inversion symmetry, and
SU(2) spin-rotation symmetry. We have shown that when spin-rotation symmetry is
preserved, the Dirac line is protected by either reflection symmetry or the 
product of inversion with time-reversal symmetry $IT$.
In the former case, the nodal lines are classified by an $M \mathbbm{Z}$ invariant~\cite{ChiuSchnyder14}, 
which takes the form of a mirror number, see Eq.~\eqref{eq_mirror_invariant}.
In the latter case the stability of the Dirac line is guaranteed
by a quantized nonzero Berry phase, which leads to a 
$\mathbb{Z}_2$ classification, see Eq.~\eqref{eq_berry_phase}.
As a representative example of a line node semimetal, we have considered Ca$_3$P$_2$~\cite{schoopZrSiS_arXiv15},
which exhibits a topologically stable Dirac ring at the Fermi energy. By means of a tight-binding model
derived from ab initio DFT calculations, we have computed the mirror number and 
the quantized Berry phase for this material (Fig.~\ref{mm_Fig3}) and shown that the Dirac band
crossing is protected by reflection or $IT$ symmetry.
The band topology of this Dirac line semimetal was also discussed
in terms of a low-energy effective theory, see Eq.~\eqref{eq_low_energy_spinless}. 

Even though the mirror invariant~\eqref{eq_mirror_invariant} does not directly give rise 
to topological surface states, Dirac line semimetals generically exhibit drumhead surface states
which are due to a quantized Berry phase. 
By deriving a relation between the mirror number~\eqref{eq_mirror_invariant}
and the Berry phase~\eqref{eq_berry_phase}, we have established a direct connection between 
the existence of   drumhead surface states and the topological stability of  Dirac nodal lines
in the bulk, see Eq.~\eqref{relation_MZ_berry}. Using the ab initio derived tight-binding model, we have computed
the surface spectrum of Ca$_3$P$_2$, showing that its drumhead surface state is weakly dispersing with an
effective mass $m^{*} \simeq  4 m_e$ [Fig.~\eqref{mm_Fig3}(b) and~\eqref{mm_Fig3}(d)].

 In Ca$_3$P$_2$ spin-rotation symmetry is conserved to a very good approximation, since
 spin orbit coupling for the light elements Ca and P is very small. However, there are   nodal line semimetals
with  heavy atoms, such as PbTaSe$_2$ and TlTaSe$_2$, in which spin-rotation symmetry is broken, 
due to the non-negligible spin-orbit interactions. In these systems the four-fold degenerate
Dirac  rings are unstable. Two-fold degenerate Weyl rings, on the other hand, can be protected
against gap opening by reflection symmetry, provided either time-reversal or reflection symmetry is broken. We have shown that the stability 
 of these Weyl rings is guaranteed by the mirror invariant~\eqref{eq_mirror_invariant}. 
 Similar to the Dirac nodal line semimetals, Weyl ring semimetals support drumhead surface
 states (Figs.~\ref{mm_Fig5} and~\ref{mm_Fig6}). The region in the surface Brillouin zone where these drumhead states appear are bounded  by the projected Weyl rings. 
 
 %On the other hand, the Berry phase integrated along the reflection direction is quantized by reflection symmetry or \emph{IT} symmetry.
%This relation provides an alternative method to compute the Berry phase by using reflection symmetry. 

Determining the stability of the drumhead surface states against disorder and interactions
needs a careful analysis of different types of scattering and interaction processes, involving both 
states near the bulk line nodes and surface states.
The drumhead surface state of Ca$_3$P$_2$ has a relatively weak dispersion (Fig.~\ref{mm_Fig3}),
which gives rise to a large density of states thereby enhancing interaction effects.
Therefore, even small interactions 
may lead to unusual symmetry-broken states at the surface, such as
surface superconductivity~\cite{KopninVolovikPRB11,tangFu_natPhys14} 
or surface magnetism~\cite{graphene_edge_magnetism}.
Disorder scattering, on the other hand, breaks the crystalline symmetries that protect
the surface states. Moreover, it mixes bulk and surface states, since there is no full gap in the bulk energy spectrum.
For the case of crystalline topological insulators it has been shown that the surface states are
robust against disorder when the disorder respects the crystal symmetries on average~\cite{diez_NJP_15}.
In appendix~\ref{appendix_disorder}, we study this question in terms of a one-dimensional reflection symmetric toy model
with a quantized Berry phase. In order to infer how impurity scattering affects the topological properties,
we determine the charge that is accumulated at the two ends of this one-dimensional system~\cite{Nodal_line_response_Taylor}.
We find that even in the presence of disorder that respects reflection symmetry on average, the end charges 
remain to a good approximation quantized to $\pm e/2$. 
Due to Eq.~\eqref{eq_end_charge_berry}, which relates the end charges to the Berry phase, this
 indicates that the bulk topological properties
remain unaffected by this type of disorder.
%When the average of disorder respects to inversion symmetry and their strength is not strong, $\pm e/2$ survives on each end. The charge accumulation directly shows that boundary modes survives under this type of the disorder. 
%If disorder respects to system symmetries and are not strong enough to cause quantum phase transition, the topological properties should be intact under the symmetry protection. 
This finding suggest that the drumhead surface states of nodal line semimetals
are not destroyed by impurities, 
as long as the disorder respects reflection symmetry on average and its strength is smaller than the
energy gap between the conduction and valence bands.
% Bulk gapless property under this type of the disorders is unknown so we leave this for our future direction. 
	
In conclusion, Dirac ring and Weyl ring semimetals are a new type of topological material which is characterized 
by a non-zero mirror invariant and a quantized Berry phase. 
The nontrivial band topology of these  systems manifests itself at the surface 
in terms of protected drumhead surface states. There are many interesting avenues for future research
on  line node semimetals. For example, the drumhead  states may give rise to
 unusual correlation physics at the surface.
Another promising direction for future work
is the study of novel topological response phenomena in these systems.

\emph{Note added}. --- Upon completion of this work, we became aware of a study by Yamakage \textit{et al}.~\cite{yamakageArXiv15}, 
which discusses the topology of line node semi-metals in terms of a $k$-independent reflection operator,
using a $k$-dependent gauge transformation. 

%Anomalous magneto-transport? Negative magnetoresistence?

 %%%%%%%%%%%%%%%%%%%
 %%%%%%%%%%%%%%%%%%%

\acknowledgments
The authors thank M.~Franz, G.~Bian, T.~Heikkil\"a, T.~Hyart, L.~Schoop, and D.-H. Xu
for useful discussions. 
The support of the Max-Planck-UBC Centre for Quantum Materials is gratefully acknowledged.
APS was supported in part by the National Science Foundation under
Grant No.\ NSF PHY11-25915. CCK was also supported by Microsoft and LPS-MPO-CMTC during the last  stage of this work.
YHC is supported by a Thematic Project at Academia Sinica. MYC acknowledges the support by the U.S. Department of Energy, Office of Science, Office of Basic Energy Sciences, under Award No. DE-FG02-97ER45632.

%%%% APPENDIX

\appendix

%%%%%%%%%%%%%%%%%%%%%%%%%%%%%%%%%%%%%%%%%%
%%%%%%%%%%%%%%%%%%%%%%%%%%%%%%%%%%%%%%%%%%

\section{Details of the tight-binding model}
\label{appencix_A}

In this Appendix we   
give a detailed description of the tight-binding 
Hamiltonian of Sec.\ref{sec_tight-binding}.

\subsection{Matrix elements}
\label{appencix_matrix_elements}

The matrix elements given below closely follows Eq.~\ref{eq:TBelements}.
The position vectors ${\bf s}_\alpha$ of each orbital are listed in Table ~\ref{table:pos}.
We illustrate each hopping terms in Fig.~\ref{Def-hop}.

\begin{table}[t]
\caption{ \label{table_postion_vectors}
Position vectors ${\bf s}_\alpha$ of each orbital. All vectors are given in the crystal coordinate system, which 
is indicated by the red/green arrows in Fig.~\ref{Def-hop}. The lattice vectors are ${\bf a}$ = [7.150, -4.218, 0.000], ${\bf b}$ = [0.000, 8.256, 0.000], and ${\bf c}$ = [0.000, 0.000, 6.836] in the unit of \r{A}.}
\centering
\setlength{\extrarowheight}{2pt}
\begin{tabular}{c *2{C{1.5in}}}
\hline\hline %inserts double horizontal lines
$\alpha$ & Orbital & ${\bf s}_{\alpha} $  \\ [1ex] % inserts table 
%heading
\hline % inserts single horizontal line
1 & Ca1 & (  0.2029 ,     0.0 ,    0.25 )   \\ % inserting body of the table
2 & Ca2 & ( -0.2029 , -0.2029 ,    0.25 )   \\
3 & Ca3 & (     0.0 ,  0.2029 ,    0.25 )   \\
4 & Ca4 & ( -0.2029 ,     0.0 ,   -0.25 )  \\
5 & Ca5 & (  0.2029 ,  0.2029 ,   -0.25 )   \\
6 & Ca6 & (     0.0 , -0.2029 ,   -0.25 )   \\
7 & P1 &  ( 0.6215 ,     0.0 ,    0.25 )   \\ 
8 & P2 & ( -0.6215 , -0.6215 ,    0.25 )   \\
9 & P3 & (     0.0 ,  0.6215 ,    0.25 )   \\
10 & P4 & ( -0.6215 ,     0.0 ,   -0.25 )  \\
11 & P5 & (  0.6215 ,  0.6215 ,   -0.25 )   \\
12 & P6 & (     0.0 , -0.6215 ,   -0.25 )   \\
\hline
\end{tabular}
\\[1.5pt]
\label{table:pos} % is used to refer this table in the text
\end{table}

\subsubsection{Ca-Ca matrix elements}

In the $H_{\mathrm Ca}$ block, we can further divide orbitals in each atomic species into those belong to the lower layer and the upper layer,
\begin{eqnarray}
H_{\mathrm Ca}
=
\begin{pmatrix}
H_{\mathrm Ca}^{ll} & H_{\mathrm Ca}^{lu} \cr
H_{\mathrm Ca}^{lu}{\dag} & H_{\mathrm Ca}^{uu} \cr
\end{pmatrix}
,
\end{eqnarray}
where sub-blocks $H_{\mathrm Ca}^{ll}$, $H_{\mathrm Ca}^{uu}$, and $H_{\mathrm Ca}^{lu}$ are $3\times3$ matrices.

\begin{figure}[tb]
\begin{center}
\includegraphics[width=0.95\columnwidth]{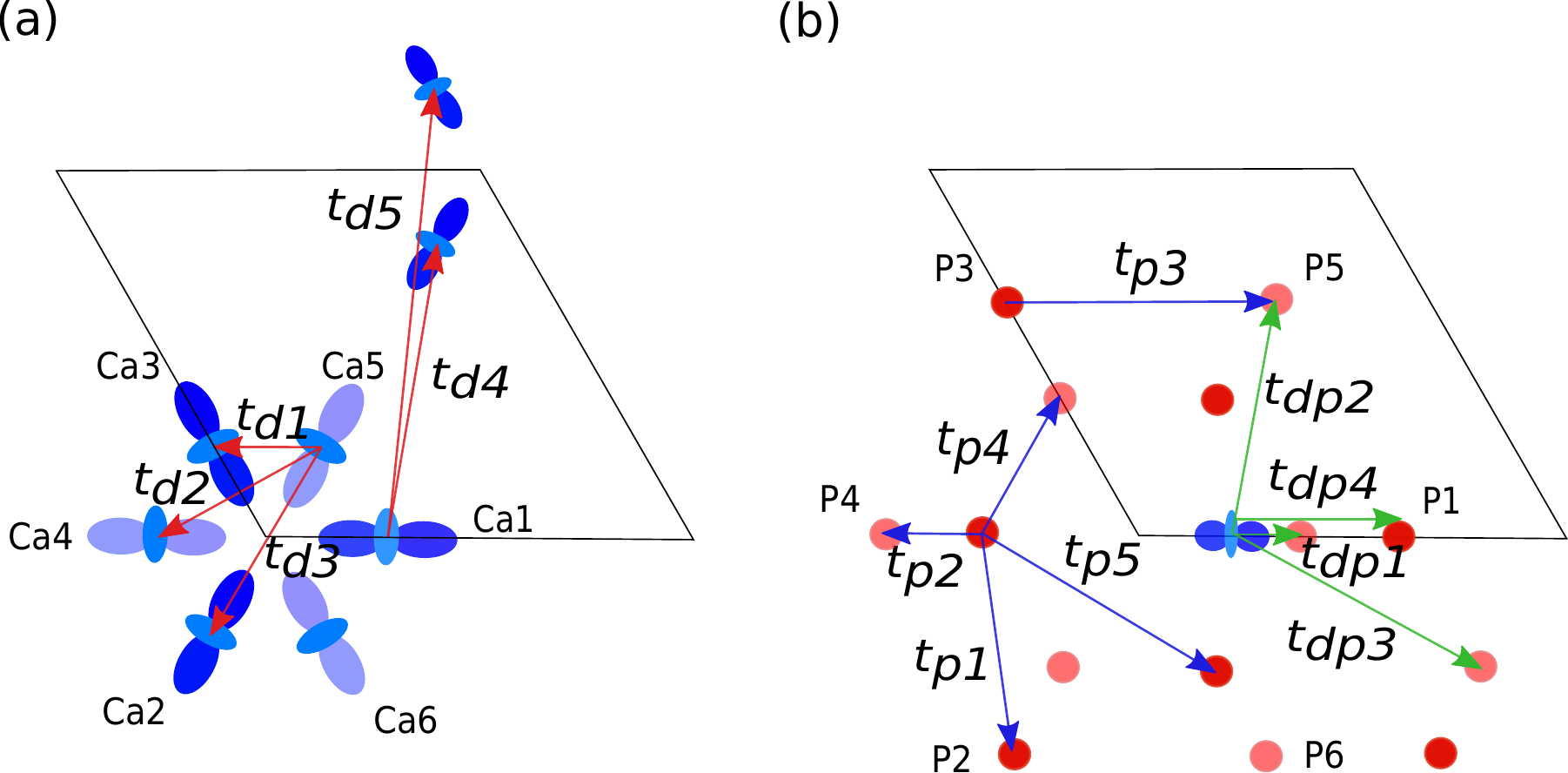}
\end{center}
  \caption{(a) Definitions of hopping integrals between two Ca orbitals. (b) Definitions of hopping integrals between two P orbitals and one Ca and one P orbital. Orbitals in the first Brillouin zone are labeled. Dark blue and red color represent orbitals in the lower plane while light blue and pink orbitals lies in the upper plane. }   
\label{Def-hop} 
\end{figure}

The Hamiltonian matrix $H_{\mathrm Ca}^{ll}$ and $H_{\mathrm Ca}^{uu}$ have 3 independent intra-layer hopping terms, the nearest-neighbor, second nearest-neighbor, and third nearest neighbor hoppings, $td_{2}$, $td_{4}$, and $td_{5}$ as shown in Fig.~\ref{Def-hop}.

\begin{eqnarray}
H_{\mathrm Ca}^{ll}
=
\begin{pmatrix}
h^{c,ll}_{11} & h^{c,ll}_{12} & h^{c,ll}_{13}\cr
h^{c,ll}_{21} & h^{c,ll}_{22} & h^{c,ll}_{23}\cr
h^{c,ll}_{31} & h^{c,ll}_{32} & h^{c,ll}_{33}\cr
\end{pmatrix}
,
\end{eqnarray}
where 
\begin{eqnarray}
h^{c,ll}_{12} = e^{i\bk \cdot \bs_{1,2}}(td_2+td_4c^4_{12}+td_5c^5_{12})\\
h^{c,ll}_{13} = e^{i\bk \cdot \bs_{1,3}}(td_2+td_4c^4_{13}+td_5c^5_{13})\\
h^{c,ll}_{23} = e^{i\bk \cdot \bs_{2,3}}(td_2+td_4c^4_{23}+td_5c^5_{23})
\end{eqnarray}
and $h^{c,ll}_{11}=h^{c,ll}_{22}=h^{c,ll}_{33}=\mu_d$.
We define phase factors $c^i_{\alpha \beta}$ for hopping integral $td_i$ with matrix indices $\alpha$ and $\beta$
\begin{eqnarray}
c^4_{12}=&e^{i\bk \cdot \bR_{100}}+e^{i\bk \cdot \bR_{110}} \\
c^4_{13}=&e^{i\bk \cdot \bR_{100}}+e^{i\bk \cdot \bR_{0-10}} \\
c^4_{23}=&e^{i\bk \cdot \bR_{-1-10}}+e^{i\bk \cdot \bR_{0-10}} \\
c^5_{12}=&e^{i\bk \cdot \bR_{010}}+e^{i\bk \cdot \bR_{0-10}} \\
c^5_{13}=&e^{i\bk \cdot \bR_{110}}+e^{i\bk \cdot \bR_{-1-10}} \\
c^5_{23}=&e^{i\bk \cdot \bR_{-100}}+e^{i\bk \cdot \bR_{100}}
\end{eqnarray}
where $\bR_{ijk}$ is the lattice vector connecting the unit cell in the $(i,j,k)$ direction and $\bs_{l,m}=\bs_m-\bs_l$.
$H_{\mathrm Ca}^{uu}$ is defined similarly.

%%%%%%%%%%%%%%%%%%%%%%%%%%%%%%%%
%%%%%%%%%%%%%%%%%%%%%%%%%%%%%%%%

$H_{\mathrm Ca}^{lu}$ contains 2 independent inter-plane hopping integrals $td_{1}$ and $td_{3}$.
\begin{eqnarray}
H_{\mathrm Ca}^{lu}
= c_0
\begin{pmatrix}
td_3e^{i\bk \cdot \bs_{1,4}} & td_1e^{i\bk \cdot \bs_{1,5}} & td_1e^{i\bk \cdot \bs_{1,6}}\cr
td_1e^{i\bk \cdot \bs_{2,4}} & td_3e^{i\bk \cdot \bs_{2,5}} & td_1e^{i\bk \cdot \bs_{2,6}}\cr
td_1e^{i\bk \cdot \bs_{3,4}} & td_1e^{i\bk \cdot \bs_{3,5}} & td_3e^{i\bk \cdot \bs_{3,6}}\cr
\end{pmatrix}
,
\end{eqnarray}
where $c_0$ is $(1+e^{i\bk \cdot \bR_{001}})$.
%
%\begin{eqnarray}
%h^{c,lu}_{12}=&td_1e^{i\tau_{15}}(1+e^{ikR_{001}})\\
%h^{c,lu}_{13}=&td_1e^{i\tau_{16}}(1+e^{ikR_{001}})\\
%h^{c,lu}_{21}=&td_1e^{i\tau_{24}}(1+e^{ikR_{001}})\\
%h^{c,lu}_{23}=&td_1e^{i\tau_{26}}(1+e^{ikR_{001}})\\
%h^{c,lu}_{31}=&td_1e^{i\tau_{34}}(1+e^{ikR_{001}})\\
%h^{c,lu}_{32}=&td_1e^{i\tau_{35}}(1+e^{ikR_{001}}),
%\end{eqnarray}
%and
%\begin{eqnarray}
%h^{c,lu}_{11}=&\,td_3e^{i\tau_{14}}(1+e^{ikR_{001}})\\
%h^{c,lu}_{22}=&\,td_3e^{i\tau_{25}}(1+e^{ikR_{001}})\\
%h^{c,lu}_{33}=&\,td_3e^{i\tau_{36}}(1+e^{ikR_{001}}).
%\end{eqnarray}

\subsubsection{P-P matrix elements}

We apply similar division of layer indices for $H_{\mathrm P}$ matrix.
\begin{eqnarray}
H_{\mathrm P}
=
\begin{pmatrix}
H_{\mathrm P}^{ll} & H_{\mathrm P}^{lu} \cr
H_{\mathrm P}^{lu}{\dag} & H_{\mathrm P}^{uu} \cr
\end{pmatrix}.
\end{eqnarray}
The Hamiltonian matrix $H_{\mathrm P}^{ll}$ and $H_{\mathrm P}^{uu}$ have 2 independent hopping integrals $tp_{1}$ and $tp_{5}$ coupling orbitals in the same layer.

\begin{eqnarray}
H_{\mathrm P}^{ll}
=
\begin{pmatrix}
h^{p,ll}_{11} & h^{p,ll}_{12} & h^{p,ll}_{13}\cr
h^{p,ll}_{21} & h^{p,ll}_{22} & h^{p,ll}_{23}\cr
h^{p,ll}_{31} & h^{p,ll}_{32} & h^{p,ll}_{33}\cr
\end{pmatrix}
,
\end{eqnarray}
where 
\begin{eqnarray}
h^{p,ll}_{12} &=& e^{i\bk \cdot \bs_{7,8}}(tp_5+tp_1a^1_{12})\\
h^{p,ll}_{13} &=& e^{i\bk \cdot \bs_{7,9}}(tp_5+tp_1a^1_{13})\\
h^{p,ll}_{23} &=& e^{i\bk \cdot \bs_{8,9}}(tp_5+tp_1a^1_{23})
\end{eqnarray}
and $h^{p,ll}_{11}=h^{p,ll}_{22}=h^{p,ll}_{33}=\mu_p$.
$a^i_{\alpha \beta}$ are phase factors from hopping $tp_i$ with matrix index $\alpha$ and $\beta$,
\begin{eqnarray}
a^1_{12}&=&e^{i\bk \cdot \bR_{100}}+e^{i\bk \cdot \bR_{110}} \\
a^1_{13}&=&e^{i\bk \cdot \bR_{100}}+e^{i\bk \cdot \bR_{0-10}} \\
a^1_{23}&=&e^{i\bk \cdot \bR_{-1-10}}+e^{i\bk \cdot \bR_{0-10}}.
\end{eqnarray}
$H_{\mathrm P}^{uu}$ can be defined similarly.

$H_{\mathrm P}^{lu}$ contains 3 independent inter-plane hopping integrals $tp_{2}$, $tp_{3}$, and $tp_{4}$.
\begin{eqnarray}
H_{\mathrm P}^{lu}
=
\begin{pmatrix}
h^{p,lu}_{11} & h^{p,lu}_{12} & h^{p,lu}_{13}\cr
h^{p,lu}_{21} & h^{p,lu}_{22} & h^{p,lu}_{23}\cr
h^{p,lu}_{31} & h^{p,lu}_{32} & h^{p,lu}_{33}\cr
\end{pmatrix}
,
\end{eqnarray}
where
\begin{eqnarray}
h^{p,lu}_{11}&=&\,tp_2e^{i\bk \cdot \bs_{7,10}}(e^{i\bk \cdot \bR_{100}}+e^{i\bk \cdot \bR_{101}})\\
h^{p,lu}_{22}&=&\,tp_2e^{i\bk \cdot \bs_{8,11}}(e^{i\bk \cdot \bR_{-1-10}}+e^{i\bk \cdot \bR_{-1-11}})\\
h^{p,lu}_{33}&=&\,tp_2e^{i\bk \cdot \bs_{9,12}}(e^{i\bk \cdot \bR_{010}}+e^{i\bk \cdot \bR_{011}}),
\end{eqnarray}
and
\begin{eqnarray}
h^{p,lu}_{12}&=&\,e^{i\bk \cdot \bs_{7,11}}(tp_3 c_0+tp_4a^4_{12})\\
h^{p,lu}_{13}&=&\,e^{i\bk \cdot \bs_{7,12}}(tp_3 c_0+tp_4a^4_{13})\\
h^{p,lu}_{21}&=&\,e^{i\bk \cdot \bs_{8,10}}(tp_3 c_0+tp_4a^4_{21})\\
h^{p,lu}_{23}&=&\,e^{i\bk \cdot \bs_{8,12}}(tp_3 c_0+tp_4a^4_{23})\\
h^{p,lu}_{31}&=&\,e^{i\bk \cdot \bs_{9,10}}(tp_3 c_0+tp_4a^4_{31})\\
h^{p,lu}_{32}&=&\,e^{i\bk \cdot \bs_{9,11}}(tp_3 c_0+tp_4a^4_{32}).
\end{eqnarray}
The corresponding phase factors are,
\begin{eqnarray}
c_0            &=& 1+e^{i\bk \cdot \bR_{001}}\\
a^4_{12}&=&e^{i\bk \cdot \bR_{0-10}}+e^{i\bk \cdot \bR_{0-11}}\\
a^4_{13}&=&e^{i\bk \cdot \bR_{110}}+e^{i\bk \cdot \bR_{111}}\\
a^4_{21}&=&e^{i\bk \cdot \bR_{0-10}}+e^{i\bk \cdot \bR_{0-11}}\\
a^4_{23}&=&e^{i\bk \cdot \bR_{-100}}+e^{i\bk \cdot \bR_{-101}}\\
a^4_{31}&=&e^{i\bk \cdot \bR_{110}}+e^{i\bk \cdot \bR_{111}}\\
a^4_{32}&=&e^{i\bk \cdot \bR_{-100}}+e^{i\bk \cdot \bR_{-101}}.
\end{eqnarray}

\subsubsection{Ca-P matrix elements}

Finally, the inter-orbital hopping matrix $V$ describes the hybridization between Ca and P orbitals. 
We again divide $V$ into four $3\times3$ matrices according to their layer indices,
\begin{eqnarray}
V
=
\begin{pmatrix}
V^{ll} & V^{lu} \cr
V^{ul} & V^{uu} \cr
\end{pmatrix}
\end{eqnarray}
.
The $V^{ll}$ and $V^{uu}$ blocks only have diagonal elements, which can be written down with the hopping integrals $tdp_4$,
\begin{eqnarray}
V^{ll}
= b_1\,tdp_4
\begin{pmatrix}
e^{i\bk \cdot \bs_{1,7}} & 0      &  0 \cr
0      & e^{i\bk \cdot \bs_{2,8}} &  0 \cr
0      & 0 &  e^{i\bk \cdot \bs_{3,9}} \cr
\end{pmatrix}
\end{eqnarray}
and
\begin{eqnarray}
V^{uu}
= -b_1\,tdp_4
\begin{pmatrix}
e^{i\bk \cdot \bs_{4,10}} & 0      &  0 \cr
0      & e^{i\bk \cdot \bs_{5,11}} &  0 \cr
0      & 0 &  e^{i\bk \cdot \bs_{6,12}} \cr
\end{pmatrix} , \quad
\end{eqnarray}
where the phase factor $b_1=(e^{i\bk \cdot \bR_{001}}-e^{i\bk \cdot \bR_{00-1}})$. 
We note that the minus sign in $V^{uu}$ is due to the opposite orientation of $p_x$ orbitals in the different layer.
Also due to the opposite inversion symmetry eigenvalue of the $p_x$ and the $d_{z^2}$ orbital, hopping integrals vanish if both of them lie in the same plane.
Hence, only hopping integrals from different unit cell contributes to diagonal elements.

Inter-layer coupling $tdp1$, $tdp2$, and $tdp3$ contributes to $V^{lu}$ and $V^{ul}$ matrices,
\begin{eqnarray}
V^{lu}
=
\begin{pmatrix}
V^{lu}_{11} & V^{lu}_{12}  &  V^{lu}_{13} \cr
V^{lu}_{21} & V^{lu}_{22}  &  V^{lu}_{23} \cr
V^{lu}_{31} & V^{lu}_{32}  &  V^{lu}_{33} \cr
\end{pmatrix}
,
\end{eqnarray}
where
\begin{eqnarray}
V^{lu}_{11} &=& -tdp_1e^{i\bk \cdot \bs_{1,10}}(e^{i\bk \cdot \bR_{101}}-e^{i\bk \cdot \bR_{100}}) \\
V^{lu}_{22} &=& -tdp_1e^{i\bk \cdot \bs_{2,11}}(e^{i\bk \cdot \bR_{-1-11}}-e^{i\bk \cdot \bR_{-1-10}}) \\
V^{lu}_{33} &=& -tdp_1e^{i\bk \cdot \bs_{3,12}}(e^{i\bk \cdot \bR_{011}}-e^{i\bk \cdot \bR_{010}}) .
\end{eqnarray}
and off-diagonal elements
\begin{eqnarray}
V^{lu}_{12} &=&\,e^{i\bk \cdot \bs_{1,11}}(tdp_2b_2 +tdp_3b^3_{12}) \\
V^{lu}_{13} &=&\,e^{i\bk \cdot \bs_{1,12}}(tdp_2b_2 +tdp_3b^3_{13})\\
V^{lu}_{21} &=&\,e^{i\bk \cdot \bs_{2,10}}(tdp_2b_2 +tdp_3b^3_{21})\\
V^{lu}_{23} &=&\,e^{i\bk \cdot \bs_{2,12}}(tdp_2b_2 +tdp_3b^3_{23})\\
V^{lu}_{31} &=&\,e^{i\bk \cdot \bs_{3,10}}(tdp_2b_2 +tdp_3b^3_{31})\\
V^{lu}_{32} &=&\,e^{i\bk \cdot \bs_{3,11}}(tdp_2b_2 +tdp_3b^3_{32}).
\end{eqnarray}
The phase factors are,
\begin{eqnarray}
b_2 &=& -(e^{i\bk \cdot \bR_{001}}-e^{i\bk \cdot \bR_{000}})\\
b^3_{12} &=& -(e^{i\bk \cdot \bR_{0-11}}-e^{i\bk \cdot \bR_{0-10}})\\
b^3_{13} &=& -(e^{i\bk \cdot \bR_{111}}-e^{i\bk \cdot \bR_{110}})\\
b^3_{21} &=& -(e^{i\bk \cdot \bR_{0-11}}-e^{i\bk \cdot \bR_{0-10}})\\
b^3_{23} &=& -(e^{i\bk \cdot \bR_{-101}}-e^{i\bk \cdot \bR_{-100}})\\
b^3_{31} &=& -(e^{i\bk \cdot \bR_{111}}-e^{i\bk \cdot \bR_{110}})\\
b^3_{32} &=& -(e^{i\bk \cdot \bR_{-101}}-e^{i\bk \cdot \bR_{-100}})
\end{eqnarray}
, where $b^i_{\alpha \beta}$ belongs to hopping $tdp_i$ between index $\alpha$ and $\beta$.

Similarly, we have 
\begin{eqnarray}
V^{ul}
=
\begin{pmatrix}
V^{ul}_{11} & V^{ul}_{12}  &  V^{ul}_{13} \cr
V^{ul}_{21} & V^{ul}_{22}  &  V^{ul}_{23} \cr
V^{ul}_{31} & V^{ul}_{32}  &  V^{ul}_{33} \cr
\end{pmatrix}
,
\end{eqnarray}
where
\begin{eqnarray}
V^{ul}_{11} &=& -tdp_1e^{i\bk \cdot \bs_{4,7}}(e^{i\bk \cdot \bR_{-10-1}}-e^{i\bk \cdot \bR_{-100}}) \\
V^{ul}_{22} &=& -tdp_1e^{i\bk \cdot \bs_{5,8}}(e^{i\bk \cdot \bR_{11-1}}-e^{i\bk \cdot \bR_{1110}}) \\
V^{ul}_{33} &=& -tdp_1e^{i\bk \cdot \bs_{6,9}}(e^{i\bk \cdot \bR_{0-1-1}}-e^{i\bk \cdot \bR_{0-10}}) .
\end{eqnarray}
and off-diagonal elements
\begin{eqnarray}
V^{ul}_{12} &=&\,e^{i\bk \cdot \bs_{4,8}}(tdp_2b_2 +tdp_3b^{3}_{12})^* \\
V^{ul}_{13} &=&\,e^{i\bk \cdot \bs_{4,9}}(tdp_2b_2 +tdp_3b^{3}_{13})^*\\
V^{ul}_{21} &=&\,e^{i\bk \cdot \bs_{5,7}}(tdp_2b_2 +tdp_3b^{3}_{21})^*\\
V^{ul}_{23} &=&\,e^{i\bk \cdot \bs_{5,9}}(tdp_2b_2 +tdp_3b^{3}_{23})^*\\
V^{ul}_{31} &=&\,e^{i\bk \cdot \bs_{6,7}}(tdp_2b_2 +tdp_3b^{3}_{31})^*\\
V^{ul}_{32} &=&\,e^{i\bk \cdot \bs_{6,8}}(tdp_2b_2 +tdp_3b^{3}_{32})^*,
\end{eqnarray}
where $^*$ denotes the complex conjugate.

\subsection{Tight-binding parameters}

We list parameters of the tight-binding model in the unit of eV below.
The hopping integrals between two Ca orbitals are $t_{d1}=-0.2031$, $t_{d2}=-0.6388$, $t_{d3}=-0.0786$, $t_{d4}=-0.216$, and $t_{d5}=0.0516$. 
Those between two P orbitals are $t_{p1}=-0.041$, $t_{p2}=-0.4077$, $t_{p3}=-0.0479$, $t_{p4}=-0.1067$, and $t_{p5}=0.0548$.
Finally, the hopping amplitudes between Ca and P orbitals are $t_{dp1}=0.1415$, $t_{dp2}=0.0379$, $t_{dp3}=0.0443$ and $t_{dp4}=0.0376$.
The chemical potentials are $\mu_d=2.6808$ and $\mu_p=-1.2186$ for Ca and P respectively.

%%%%%%%%%%%%%%%%%%%%%%%%%%%%%%%%%%%%%%%%%%
%%%%%%%%%%%%%%%%%%%%%%%%%%%%%%%%%%%%%%%%%%

\section{topological number and Berry phase}\label{quantized Berry} \label{appendixB}

To show that the Berry phase in the $k_z$ direction is quantized and is related to ${n^{+}_{\textrm{occ}} }$ in Eq.~\eqref{relation_MZ_berry}, we recall some basic facts of inversion symmetry. 
We assume no degeneracies so the inversion symmetry acts the wavefunctions $\ket{u_{k,j}}$ in the unique expression ($k\equiv k_z$)
\bee
\ket{u_{-k,j}}=e^{-i\alpha_k^j}R_k\ket{u_{k,j}}
\ee
The reflection operator obeys $R_{-k}R_{k}=\pm \bI$ for spinless/spin-$1/2$ systems respectively. For spin-$1/2$, we redefine $R_k \rightarrow -iR_k$ so that $R_{-k}R_{k}= \bI$. Also, $R_{k}^\dagger R_k=\bI$. Let us rewrite the Berry phase 
\begin{small}
\begin{eqnarray}
\mathcal{P}&=&-i \big ( \int_0^{\pi} + \int_{-\pi}^0 \big) \sum_{E_j< E_{\textrm{F}}}      \bra{u_{k,j}} \partial_{k} \ket{u_{k,j}} dk           \nonumber \\
&=&-i \int_{0}^{\pi}  \sum_{E_j< E_{\textrm{F}}}      \bra{u_{k,j}} \partial_{k} \ket{u_{k,j}} dk           \nonumber \\
 &\ &+i \int_{0}^{\pi}  \sum_{E_j< E_{\textrm{F}}}      \bra{u_{k,j}} R_{k}^\dagger e^{i\alpha_k^j} \partial_{k} e^{-i\alpha_k^j } R_{k}\ket{u_{k,j}} dk \nonumber \\
&=&\sum_{E_j  < E_{\textrm{F}}} (\alpha_\pi^j - \alpha_{0}^j) +  i \int_{0}^{\pi}  \sum_{E_j< E_{\textrm{F}}}      \bra{u_{k,j}} R_{k}^\dagger \partial_{k} R_{k}\ket{u_{k,j}} dk \nonumber \\ \label{Berryphase invariant}
\end{eqnarray}
\end{small}
%\begin{align}
%P_1=&-i \int_{-\pi}^{\pi}  \sum_{E_j<0}      \bra{u_{k,j}} \partial_{k} \ket{u_{k,j}} dk           \nonumber \\
%=&-i \int_{-\pi}^{\pi}  \sum_{E_j<0}      \bra{u_{-k,j}} P_{-k}^\dagger e^{-i\alpha_k^j} \partial_{k} e^{i\alpha_k^j } P_{-k}\ket{u_{-k,j}} dk \nonumber \\
%=&-P_1 + \sum_{E_j < 0} (\alpha_\pi^j - \alpha_{-\pi}^j) \nonumber  \\
%&-i \int_{-\pi}^{\pi}  \sum_{E_j<0}      \bra{u_{-k,j}} P_{-k}^\dagger \partial_{k} P_{-k}\ket{u_{-k,j}} dk 
%\end{align}
%For the trivial insulator, the Berry phase vanishes $P_1=0$ so the relation between the phases and the wavefunctions of the trivial system
%\bee
%\sum_{E'_j < 0} ({\alpha'}_\pi^j - {\alpha'}_{0}^j) 
%= -i \int_{0}^{\pi}  \sum_{E'_j<0}     \bra{u'_{k,j}} P_{k}^\dagger \partial_{k} P_{k}\ket{u'_{k,j}} dk 
%\ee
%{\cb something is absent then we have }
The reflection symmetry operator has a generic block-diagonalized from~\cite{chiu_review15}
\bee
R_k=U_{i_1j_1}e^{in_1 k }\oplus U_{i_2j_2}e^{in_2 k }\oplus\ldots \oplus U_{i_Nj_N}e^{in_N k },
\ee
where $U_{i_lj_l}$ is a unitary matrix and we use the lattice constant $a\equiv 1$. 
\bee
R_k^\dagger \partial_k R_k= in_1\delta_{i_1j_1}\oplus in_2\delta_{i_2j_2} \oplus \ldots \oplus in_N\delta_{i_Nj_N}.
\ee
Hence, $\partial R$ is just $m\pi$, where $m$ is an integer
\begin{align}
 i \int_{0}^{\pi}  \sum_{E_j< E_{\textrm{F}} }      \bra{u_{k,j}} R_{k}^\dagger \partial_{k} R_{k}\ket{u_{k,j}} dk = - \sum_{l=1}n_l m_l \pi,
\end{align}
where $m_l$ is the number of the occupied states in $U_{i_lj_l}$ block. 
% Incorrect 
%\begin{align}
%&\int_{0}^{\pi}  \sum_{E'_j<0}     \bra{u'_{k,j}} P_{k}^\dagger \partial_{k} P_{k}\ket{u'_{k,j}} dk \nonumber \\
%=& \int_{0}^{\pi}  \sum_{E_j<0}      \bra{u_{k,j}} P_{k}^\dagger \partial_{k} P_{k}\ket{u_{k,j}} dk 
%\end{align}
%The Berry phase is written in the form of extra phases from reflection 
%\bee
%P_1=\sum_{E_j< 0}(\alpha_\pi^j - \alpha_{0}^j -{\alpha'}_\pi^j + {\alpha'}_{0}^j ) \label{Berryphase invariant}
%\ee
Consider left hand side of Eq.~\eqref{relation_MZ_berry}
\begin{align}
n^{+,\pi}_{\textrm{occ}} - n^{+,0}_{\textrm{occ}}= & \frac{1}{2}\sum_{E_j < 0} \big ( \bra{u_{\pi,j}} R_\pi \ket{u_{\pi,j}} -  \bra{u_{0,j}} R_0 \ket{u_{0,j}}  \big ) \nonumber \\
=& \frac{1}{2} \sum_{E_j <0 } (e^{i \alpha_\pi^j}-e^{i\alpha_0 ^j})
\end{align}
Since $R_{k_0}^\dagger=R_{k_0}$, where $k_0=-k_0$, such as $0,\ \pi$ so $e^{i\alpha_{k_0}^j}=\pm 1$ and then  
\bee
n^{+,\pi}_{\textrm{occ}} - n^{+,0}_{\textrm{occ}}\equiv  \frac{1}{\pi}\sum_{E_j < E_{\textrm{F}}} (\alpha_\pi^j - \alpha_0^j) \ (\rm{mod}\ 2) \label{mod2}
\ee
Thus, $(-1)^{n^{+,\pi}_{\textrm{occ}} - n^{+,0}_{\textrm{occ}}}=e^{i \sum_{E_j< E_{\textrm{F}} } (\alpha_\pi^j - \alpha_0^j) }$. By using Eq.~\eqref{Berryphase invariant},~~\eqref{mod2}, we obtain the relation in Eq.~\eqref{relation_MZ_berry} between the topological invariants and the Berry phase $\mathcal{P}$ is either $0$ or $\pi$ (mod $2\pi$) since $2n\pi$ phase can be cancelled  by a large $U(1)$ gauge transformation. 

\begin{center}
*****
\end{center}

Similarly, \emph{IT} symmetry, the composite symmetry of time-reversal and inversion, also quantizes the Berry phase when $d\bk$ is integrated along any closed loop. Since time-reversal and inversion operators both flip $\bk$, the composite symmetry operators keep the same $\bk$. The integration path can be arbitrarily chose to preserve \emph{IT} symmetry. Unlike the Berry phase under reflection symmetry, the integration path should be strictly in the $k_z$ direction to preserve reflection symmetry. 

	\emph{IT} symmetry operator is the combination of a unitary matrix and complex conjugation $T\mathcal{I}=U K$; the unitary matrix $U$ might be $\bk$-dependent. To simplify the problem, we assume $U$ is $\bk$-independent, which is the case of Ca$_3$P$_2$ tight-binding model. The relation of wavefunctions under \emph{IT} symmetry is given by 
\bee
\ket{u_{\bk,j}}=e^{i\beta_{\bk}^j}U\ket{u_{\bk,l_j}}
\ee
We note that $\ket{u_{\bk,j}}$ and $\ket{u_{\bk,l_j}}$ in the same energy level might be orthogonal or identical. Let us show the Berry phase is quantized
\begin{align}
\mathcal{P}=&-i \oint  \sum_{E_j<E_{\textrm{F}} }      \bra{u_{\bk,j}} \partial_{\bk} \ket{u_{\bk,j}} d\bk  \nonumber \\
=& -i \oint  \sum_{E_j<E_{\textrm{F}} }      \bra{u_{\bk,l_j}^*} U^\dagger e^{-i\beta^j_\bk} \partial_{\bk} e^{i\beta^j_{\bk}} U \ket{u_{\bk,l_j}^*} d\bk \nonumber \\
=&  \sum_{E_j<E_{\textrm{F}} }  (\beta^j_{+}-\beta^j_{-})-i \oint  \sum_{E_j<E_{\textrm{F}}}      \bra{u_{\bk,l_j}^*} \partial_{\bk}  \ket{u_{\bk,l_j}^*} d\bk,
\end{align}
where $\beta^j_{\mp}$ represent the phases at the beginning and end of the integration path respectively. The first summation is $2n\pi $. Since the $j$th and the $l_j$th states share the same energy and each state in the second summation should be orthogonal, we safely change the index $l_j$ to $j$ in the summation. We use the identity 
\bee
 \bra{u_{\bk,j}^*} \partial_{\bk}  \ket{u_{\bk,j}^*}= \braket{\partial_{\bk} u_{\bk,j}}{u_{\bk,j}}=- \bra{u_{\bk,j}} \partial_{\bk}  \ket{u_{\bk,j}}
\ee
The Berry phase is quantized 
\bee
\mathcal{P}= \sum_{E_j<E_{\textrm{F}}}  (\beta^j_{+}-\beta^j_{-})  = n\pi
\ee

%%%%%%%%%%%%%%%%%%%%%%%%%%%%%%%%%%%%%%%%%%
%%%%%%%%%%%%%%%%%%%%%%%%%%%%%%%%%%%%%%%%%%

\section{Toy model of topological nodal lines}\label{toy lattice}
The tight-binding model of Ca$_3$P$_2$ provides the way to investigate topological nodal lines in a realistic model. However, to capture the physical features of the nodal lines only the low energy theory is needed. 
We extend the low energy theory to a simple lattice model in order to provide an economic way to investigate topological nodal lines.  Although the space group of Ca$_3$P$_2$ is $P6_3/mcm$, we consider square lattice and extend and transfer the low energy equation (\ref{eq_low_energy_spinless}) with spins to the lattice form 
\begin{align}
H_{\text{spinful}}^{\text{lattice}}({\bf k}) &= \frac{\nu'_\parallel}{a^2}g(k_\parallel)\tau_z \sigma_0 
+\frac{\nu_z}{c} \sin c k_z \tau_y \sigma_0 \nonumber \\
& + \big ( \frac{\nu'_0}{a^2} g(k_\parallel)+ V_0 \big ) \tau_0 \sigma_0  + H_{\cos k_z}
\end{align}
where $g(k_\parallel)=1+\cos a k_0 - \cos ak_x -\cos a k_y$, the lattice constants $a=8.26$~\AA and $c=6.84$~\AA, $\nu_\parallel'=\frac{2\nu_\parallel ak_0}{\sin ak_0}$, and $\nu_0'=\frac{2\nu_0 a k_0}{\sin a k_0}$.  Furthermore, we define 
\bee
H_{cos k_z}=(1-\cos c k_z )\big ( Z_\tau\tau_z \sigma_0+ Z_0 \tau_0 \sigma_0  \big )
\ee
in the simplest form so that the Berry phase inside the nodal ring is nonzero when the spin degree of freedom is neglected. By fitting the energy spectrum from the DFT calculation as $k_z=0,\ \pi$, we have $Z_\tau=0.287$~eV and $Z_0=-0.156$~eV.

%The low energy 
%\begin{align}
%H_{\text{spinful}}({\bf k})=\nu_\parallel(k_{\parallel}^2-k_0^2)\tau_z\sigma_0+\nu_z k_z\tau_y\sigma_0 + f({\bf k})\bm{1}
%\end{align}
%where $f({\bf k})=\nu_0(k_\parallel^2- k_0^2)+V_0$.   

\section{Quantized end charge under disorders}
\label{appendix_disorder}

%%%%%%%%%%%%%%%%%%
\begin{figure}[tb]
\begin{center}
\includegraphics[clip,width=0.98\columnwidth]{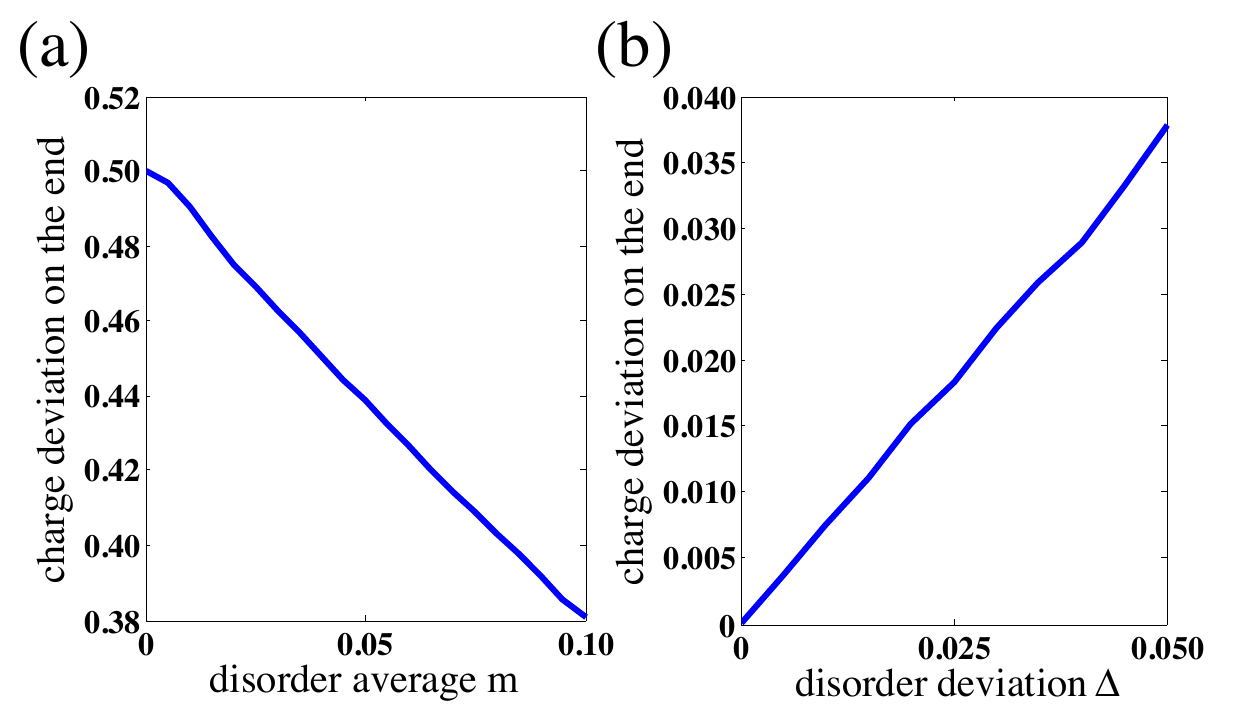} 
\end{center}
  \caption{The numerical result for the 1d inversion-symmetric topological insulator with 200 sites. We consider the half-filling scenario and compute the absolute value of charge accumulated on the first 10 sites under gaussian disorder as fixed disorder average $m$ and disorder random deviation $\Delta$ 3000 times. (a) As $\Delta=0.02$, the end charge is not quantized when the average disorder is not zero.  In the special condition that the average disorder vanishes so inversion symmetry on average is preserved, the end charge on average is $\pm e/2$. (b) The standard deviation of the disorder grows as the deviation of the end disorder grows. }   \label{disorder_fig}
\end{figure}
%%%%%%%%%%%%%%%%%%%

To understand the robustness of the topology under disorder we consider the toy model of a 1d inversion-symmetric topological insulator. We note that in a 1d system inversion symmetry is equivalent to reflection symmetry; reflection symmetric nodal lines with fixed $k_x, k_y$ is equivalent to the 1d inversion symmetric topological insulator; the Berry phase, which is the integration along the 1d BZ, is quantized. The toy model in momentum space can be simply written as 
\bee
H(p)=(\mu+\cos p)\sigma_x + \sin p \sigma_y+ \delta \cos p \bm{1} ,
\ee
which preserves inversion symmetry by satisfying Eq.~\eqref{eq_Inversion_eq} with inversion symmetry operator $I=\six$. Broken chiral symmetry caused by $\delta \cos p \bm{1}$ destroys the definition of winding number so the Berry phase is the only valid topological invariant. Furthermore, by Eq.~\eqref{def_TRsym} time-reversal symmetry is preserved with time-reversal operator $T=K$. \emph{IT} symmetry also guarantees the quantized Berry phase. By choosing $\mu=0.5$ and $\delta=0.1$, the Berry phase $\mathcal{P}=\pi$ leads to the presence of charge $\pm e/2$ at each end, which is one of the topological features of this inversion symmetric insulator. The sign of the charge depends on the occupation of the end mode. Hence, we can numerically compute the charge on one of the ends. If the charge is no longer $\pm e/2$ under disorder, the topology is destroyed by disorders.

	We add inversion symmetry breaking disorder $r_jc_j^\dagger \siz c_j^{}$ to the Hamiltonian in real space 
\bee
\hat{H}=\sum_j \big [  \frac{\mu}{2} c^\dagger_j \six c_j^{}  + c^\dagger_{j+1} \frac{\six + \delta \bm{1} +i \siy}{2} c_j +  \rm{h.c.} ],
\ee
where $r_j$ is a random number from $-\Delta+m$ to $\Delta + m$. When $m=0$, the average $<r_j>=0$ indicates the average disorder preserves inversion symmetry. As shown in Eq.~\eqref{disorder_fig} (a) when $m=0$, the charge on one end is $\pm e/2$ on average. When inversion symmetry is broken on average, the charge is no longer quantized and then the topological phase is destroyed. Fig.\ \ref{disorder_fig} (b) the standard deviation of the disorder is proportional to the deviation of the end disorder. Thus, the quantized end charges survive when disorder on average is zero and the fluctuation is small enough.

%%%%%%%%%%%%%%%%%
%%%%%%%%%%%%%%%%%%%

%%%% REFS

\bibliographystyle{apsrev4-1}
\bibliography{Ca3P2}

\begin{thebibliography}{81}
\bibitem [{foo()}]{footnote1}%
  \BibitemOpen
  \href@noop {} {}\bibinfo {note} {This relation also holds if there is spin
  degeneracy, in which case $\partial R$ is zero since the two spin components
  give opposite contribution.}\BibitemShut {Stop}%
\end{thebibliography}

\appendix

\end{document}